# Robust topology optimization of structures under uncertain propagation of imprecise stochastic-based uncertain field

Kang Gao [a, b], Duy Minh Do [c], Sheng Chu [b], Gang Wu [a], H. Alicia Kim [d], Carol A. Featherston [b*]

[a] Key Laboratory of Concrete and Prestressed Concrete Structures of the Ministry of Education, School of Civil Engineering, Southeast University, Nanjing 210096, China

[b] School of Engineering, Cardiff University, The Parade, Cardiff CF24 3AA, UK

[c] Department of Civil Engineering, Monash University, Australia

[d] Structural Engineering Department, University of California San Diego, CA 92093, USA

**Abstract**

This study introduces a novel computational framework for Robust Topology Optimization (RTO) considering imprecise random field parameters. Unlike the worst-case approach, the present method provides upper and lower bounds for the mean and standard deviation of compliance as well as the optimized topological layouts of a structure for various scenarios. In the proposed approach, the imprecise random field variables are determined utilizing parameterized *p*-boxes with different confidence intervals. The Karhunen-Loève (K-L) expansion is extended to provide a spectral description of the imprecise random field. The linear superposition method in conjunction with a linear combination of orthogonal functions is employed to obtain explicit mathematical expressions for the first and second order statistical moments of the structural compliance. Then, an interval sensitivity analysis is carried out, applying the Orthogonal Similarity Transformation (OST) method with the boundaries of each of the intermediate variable searched efficiently at every iteration using a Combinatorial Approach (CA). Finally, the validity, accuracy, and applicability of the work are rigorously checked by comparing the outputs of the proposed approach with those obtained using the particle swarm optimization (PSO) and Quasi-Monte-Carlo Simulation (QMCS) methods. Three different numerical examples with imprecise random field loads are presented to show the effectiveness and feasibility of the study.

*Corresponding author: featherstonca@cardiff.ac.uk

# 1. Introduction

At the early stages of design, topology optimization (TO) offers an opportunity to achieve creative structural configurations by pushing the material distribution to the limit within prescribed design constraints. In comparison to conventional shape and size optimization, topology optimization has a more powerful capability to determine the optimal material layout without depending on the designer's prior knowledge. Inspired by Bendsøe and Kikuchi[1], a number of methods have been successfully developed over the last two decades, including solid isotropic material with penalization (SIMP)[2, 3], evolutionary structural optimization (ESO) [4, 5] and level set-based methods (LSMs)[6-9]. These methods have greatly promoted the application of TO in structural engineering (e.g., aerospace, electronics, biomedical, automotive and civil engineering) in both scientific and industrial fields. Although studies using currently available TO methods have led to significant improvements in structural design, it is noted that the results are in most cases, underpinned by deterministic assumptions based on nominal material properties and loading conditions as well as the predetermined geometries. For a more representative approach, it is essential to consider the influences of stochastic variables on structural performance to achieve robust and reliable designs.

Studies incorporating uncertainties into topology optimization have therefore, attracted increased attention in the past few years. Generally, TO problems with uncertainties can be classified into two categories: reliability-based topology optimizations (RBTO) [10-13] and robust topology optimizations (RTO) [14-20]. RBTO aims at designing for a given risk at the expected probability level and thus ensuring that circumstances that can lead to catastrophic failure are impossible. Maute and Frangopol [12] studied RBTO problems relating to micro-electro-mechanical systems by considering uncertainties in loading, boundary conditions and material properties. First order reliability analysis methods were employed to determine structural performance. Utilizing the same method, Jung and Cho [10] investigated the RBTO design of three-dimensional Mindlin plate structures with results demonstrating that RBTO exhibits better performance than both the safety factor and worst-case approaches. A non-probabilistic RBTO framework was proposed by Kang and Luo to study the effect of geometrical nonlinearities [11]. Unlike RBTO methods, RTO methods attempt to minimize the mean [16, 17] and/or the combination of mean and standard deviation [14, 15, 18-20] of the compliance, while reducing the sensitivity of the objective function with respect to the random variables simultaneously. Alvarez and Carrasco[21] demonstrated the RTO of a truss structures by considering minimum expected compliance under stochastic loadings. An efficient and accurate approach for RTO was introduced by Dunning et al. [14-16]. In this study, the mean and standard deviation of the expected compliance were derived explicitly based on probability density functions for both stochastic loading magnitudes and directions. Besides the use of RTO for static problems, its application to vibration/dynamic problems has also become popular. Zhang et al. [22] studied the dynamic compliance of structures subjected to uncertain harmonic excitations. They also [23] presented the use of RTO for structural dynamics problems considering diffuse-region width uncertainty in which random responses were represented by a polynomial chaotic expansion.

Possible sources of uncertainty in TO problems are load conditions and material properties as well as geometric variations due to observation errors, incomplete information, manufacturing defects, etc. Among all of these, loading uncertainty is the most significant for two reasons. Firstly, at the early conceptual design phase, the magnitudes, directions, and positions of external loads on the design domain cannot be specifically determined and secondly in engineering practice applied loads are likely to experience variations during service lifetime. Moreover, since these uncertain loads are directly transferred to the design boundary they will in turn, lead to variations in structural compliance. Thus, various methods have been introduced to investigate the impact of loading uncertainties on structural topology optimization. The traditional way of solving the RTO with loading uncertainties is to use a nested double-loop optimization where the outer loop is used to minimize structural compliance with respect to the design variables and the inner loop identifies the worst-case load conditions for all possible load sets. While such a method is time-consuming in terms of computation effort, Zheng et al. [24-26] proposed a semi-analytical method for solving a series of RTO problems by combining orthogonal decomposition with uniform sampling. The method was efficient as the calculation of the uncertain variables related to compliance was outside the finite element analysis. Wu et al [27] proposed the Chebyshev interval inclusion function for RTO problems with interval uncertainty for both load magnitudes and directions. In addition to the magnitude, load positions/locations may also present degrees of uncertainty as a change in loading environment, resulting in the relocation of forces on a structure will affect its response dramatically. Liu and Wen [28] introduced the cloud model to stimulate uncertainties in loading locations. A second-order Taylor series expansion was utilized by Wang and Gao [29, 30] to derive structural compliance due to load location disturbance in which load position uncertainty was represented by an uncertain-but-bounded interval variable.

Probabilistic topology optimization analysis [14, 16-18, 26, 31-34] has been widely applied by researchers for cases where complete information on the structure is accessible, and the probability models are well-established. In engineering practice, however, it is scarcely possible to obtain sufficient samples to model probability and small discrepancies may lead to large deviations in outcome. Consequently, non-probabilistic approaches (i.e., interval analysis[35-38], fuzzy set theory[39, 40], and convex modelling[11, 13]) have been recognised as alternative methods for managing randomness with insufficient data. The stochastic physical models found in engineering practices however, do not always follow clear cut 'YES–NO' logic; instead, they are often vague, imprecise and indecisive. Under such circumstances, using probabilistic or non-probabilistic models individually may not fully illustrate the random characteristics of the physical problems. In these cases the idea of imprecise randomness combining standard probabilistic analysis with uncertain-but-bounded ranges was proposed by Beer et al. [41]. This has led to the concept of the 'imprecise random field' in uncertain computational mechanics[42]. In this approach commonly utilized random variables can be replaced by continuous random fields more suitable for simulating the stochastic properties of physical models incorporating fluctuations in the spatial domain[43]. Whilst RTO problems considering random field loading or random field material properties have been conducted in [19, 33, 44, 45], the study of imprecise stochastic-based uncertain fields has to the authors' knowledge not been reported. The application of imprecise randomness for spatially varying field parameters presented here will therefore further enhance RTO problems to attain the optimal design using a more generic approach which is consistent with the level of information more commonly available in practical design.

Following this idea, a new framework for RTO problems under polymorphic uncertainty of imprecise stochastic-based uncertain fields is introduced to investigate the upper and lower bounds of the mean and standard deviation of compliance as well as the final topological layout of a structure. The spatial randomness of the external loads in this study is generated using a parameterized *p*-box with various confidence intervals. First, the traditional exponential function is employed to simulate the covariance function of the random field and a significance check is performed to select suitable truncated terms from the K-L expansion. The linear superposition method in conjunction with a linear combination of orthogonal functions is employed to obtain explicit mathematical expressions for both the mean and standard deviation of the structural compliance with respect to the imprecise random loads. Application of the combinatorial approach (CA) then allows an efficient search for the boundaries of each intermediate variable. Next an interval sensitivity analysis is carried out by applying the Orthogonal Similarity Transformation (OST) method to further improve the computational efficiency. Finally, the sensitivity is filtered using improved Heaviside filtering and updated using the globally convergent method of moving asymptotes (GCMMA) algorithm. Iterations are performed until an optimal solution is obtained when the volume constraint is satisfied and convergence is achieved. The validity and accuracy, as well as the applicability of the proposed computational framework are rigorously verified in three steps. First the use of the linear superposition method is compared with the direct method (which generates a large number of random variables) based on calculation of the compliance. Implementation of the CA is then checked by carefully examining the monotonic property of the stochastic system for the interval variables. Finally, the accuracy of the proposed RTO algorithm is examined by comparison with low-discrepancy sequences based on the high-order nonlinear particle swarm optimization (LHNPSO) approach and the Quasi-Monte-Carlo Simulation (QMCS) method, respectively.

The paper is organized as follows: Sect. 2 presents the theory and formulations for the imprecise random field for uncertainty modelling. The RTO of structures with imprecise random field loading is presented in Sect. 3. The proposed sensitivity analysis for imprecise random field loads is presented in Sect.4 and the solution procedure for the statistical responses of the present study is explained in Sect.5. Some numerical examples are demonstrated in Sect. 6 and finally, some concluding remarks are drawn in Sect. 7.

## 2. Description of the imprecise random field for uncertainty modelling

### 2.1. Imprecise random field

Generally, stochastic processes are mathematical models used to represent random quantities evolving in time or space. For example, a probability space is a triple ($\Omega$, $\mathscr{F}$, $P$), where $\Omega$ is the sample space, the $\sigma$-algebra $\mathscr{F}$ is a collection of subsets of $\mathscr{F}$ and $P$ denotes the probability measure.

Unlike a traditional stochastic process $\{X(t,\omega)\}_{t\in T}$, which represents the varying of random values over time $t$ where $T$ is the time space, random field $\{H(\boldsymbol{x},\omega)\}_{x\in D}$ takes values that are spatial coordinates $\boldsymbol{x}$ in the topological space $D \in \mathbb{R}^n$ and D belongs to the real set, which is n-dimensional Euclidean space $\mathbb{R}^n$. Therefore, a random field can be identified as a generalization of the stochastic process with the function satisfying $H(\boldsymbol{x},\omega): D \times \Omega \mapsto \mathbb{R}^n$, which mean mapping the structural domain D into space domain $\Omega$ to get the subset $\mathbb{R}^n$, with parameters $\boldsymbol{x} \in D$ and $\omega \in \Omega$. Consider an isotropic homogenous random field with mean $\mu_{H(\boldsymbol{x},\theta)} = E[H(\boldsymbol{x},\omega)]$, and autocovariance function $Cov_H(\boldsymbol{x}_1,\boldsymbol{x}_2) = E\left[\left(H(\boldsymbol{x}_1,\omega) - \mu_{H(\boldsymbol{x}_1,\omega)}\right)\left(H(\boldsymbol{x}_2,\omega) - \mu_{H(\boldsymbol{x}_2,\omega)}\right)\right]$. The mean of the random field $\mu_{H(\boldsymbol{x},\omega)}$ is a constant in the domain and the autocovariance function $Cov_H(\boldsymbol{x}_1,\boldsymbol{x}_2)$ only relies on the absolute distance between adjacent points $|\boldsymbol{x}_1 - \boldsymbol{x}_2|$ with the standard deviation of the random field $\sigma_{H(\boldsymbol{x},\omega)}$.

When the statistical information is complete and the autocovariance function is known, one can easily make a fully objective estimate for the given events. However, in engineering practice, such large, thorough data sets are often not available, as experimental activities are both time

$F_H(\mathbf{x},\omega)$, one can obtain

$$\Xi = \left\{ F_H(\mathbf{x},\omega) \middle| \forall \mathbf{x} \in \mathbb{R}^n, \underline{F}_H(\mathbf{x},\omega) \leq F_H(\mathbf{x},\omega) \leq \overline{F}_H(\mathbf{x},\omega) \right\} \qquad (1)$$

where $\underline{F}_H(\mathbf{x},\omega)$ and $\overline{F}_H(\mathbf{x},\omega)$ represent the lower and upper bounds of the Cumulative Distribution Functions (CDFs) bounding the p-box respectively. According to the concept of the parameterised *p*-box, a more general representation of the imprecise probability distribution $\Xi^p \sim \left( \left[ \underline{\mu}_H, \overline{\mu}_H \right], \left[ \underline{\sigma}_H, \overline{\sigma}_H \right] \right)$ can be formulated as

$$\Xi^p = \left\{ F_H(\mathbf{x},\omega; \mu_H, \sigma_H) \middle| \forall \mathbf{x} \in \mathbb{R}^n, \mu_H \in \left[ \underline{\mu}_H, \overline{\mu}_H \right], \sigma_H \in \left[ \underline{\sigma}_H, \overline{\sigma}_H \right] \right\} \qquad (2)$$

where $\Xi^p$ is the parameterised p-box with imprecise mean and standard deviation. $\underline{\mu}_H, \overline{\mu}_H$ are the lower and upper bounds of the mean and $\underline{\sigma}_H, \overline{\sigma}_H$ the lower and upper bounds of standard deviation respectively.

### 2.2. Discretization of the imprecise random field

As one of the key steps in modelling a random field, stochastic discretization has a direct influence on the computational efficiency and accuracy of a probability analysis. Several random field discretization methods have already been proposed. There are two categories based on the probability distribution of the random field, Gaussian and non-Gaussian random fields. Compared with non-Gaussian, the application of Gaussian random fields in stochastic analysis is much wider. This is due to the fact that Gaussian random fields have the following outstanding features: 1. They are characterized by second-order statistics; 2. They are stable in linear combinations; 3. It is easy to capture marginal and conditional distributions. Therefore, the present study only considers Gaussian random fields.

The Karhunen-Loève (K-L) expansion is an efficient technique for stochastic discretization using limited truncated terms. To represent the spectral description of the imprecise random field, the Karhunen-Loève expansion can be extended to

$$H^p(\mathbf{x},\omega) = \mu_H^I + \sum_{i=1}^{\infty} \sqrt{\lambda_i^I} \psi_i(\mathbf{x}) \xi_i(\omega) \qquad (3)$$

where $\mu_H^I = [\underline{\mu}_H, \bar{\mu}_H]$ is the imprecise mean. $\lambda_i^I \in [0, \infty)$ are the imprecise eigenvalues, $\psi_i(x): D \to \mathbb{R}^n$ are the corresponding orthogonal eigenfunctions of the autocovariance kernel and $\xi_i(\omega)$ are mutually uncorrelated random variables with zero mean and unit variance, which have the following properties:

$$E(\xi_i(\omega)) = 0 \text{ and } E(\xi_i(\omega)\xi_j(\omega)) = \delta_{ij} \tag{4}$$

where $\delta_{ij}$ is the Kronecker delta function. An explicit expression for $\xi_i(\omega)$ can be obtained from Eq.(3) by multiplying by $\psi_i(x)$ and integrating over the domain D, as follows

$$\xi_i(\omega) = \frac{1}{\sqrt{\lambda_i^I}} \int_D \left( H^p(x, \omega) - \mu_H^I \right) \psi_i(x) dx \tag{5}$$

Utilizing Mercer's theorem, the autocovariance function obeys the spectral decomposition:

$$Cov_H(x_1, x_2) = \sum_{i=0}^{\infty} \lambda_i^I \psi_i(x_1) \psi_i(x_2) \tag{6}$$

The set of eigenpairs $\{\lambda_i^I, \psi_i\}$ in Eq. (6) can be obtained using a homogeneous Fredholm integral equation of the second kind, as

$$\int_D Cov_H(x_1, x_2) \psi_i(x_1) dx_1 = \lambda_i^I \psi_i(x_2) \tag{7}$$

In practice, the random field $H^p(x, \omega)$ can be approximated by $\tilde{H}^p(x, \omega)$ after truncating the series at the $M$-th term without sacrificing too much accuracy, as

$$H^p(x, \omega) \approx \tilde{H}^p(x, \omega) = \mu_H^I + \sum_{i=1}^{M} \sqrt{\lambda_i^I} \psi_i(x) \xi_i(\omega) \tag{8}$$

where $M$ is determined by the significance check,

$$s = \sum_{i=1}^{M} \lambda_i \bigg/ \sum_{i=1}^{\infty} \lambda_i \geq s_0 \tag{9}$$

where $s_0$ is a predefined threshold value.

## 3. Robust topology optimization of structures with imprecise random field loads

*3.1. Robust topology optimization*

A density-based method is utilized to perform robust topology optimization for structures with loading uncertainty. Using this approach, the given domain is discretised into a grid of finite elements each of which is filled according to a density variable. To avoid the binary, on-off problem, a continuous density distribution function is used for each element. The modified SIMP interpolation scheme is as follows,

$$E_e = E_{min} + (E_0 - E_{min})\rho_e^P \tag{10}$$

where $\rho_e^P$ ($0 \leq \rho_e^P \leq 1$) are the intermediate densities of element *e*. $E_0$, $E_{min}$ are the Young's moduli of the solid and void phases, respectively. $E_{min}$ is a small non-zero number. Intermediate densities are penalized by power *p*. Typically, the penalization parameter *p* is chosen as 3 for the objective function.

The robust topology optimization problem of this study can be written in terms of the statistical moments of compliance

$$\begin{aligned}
\min_{\rho} : \quad & J = \mu(c(\omega)) + \beta\sigma(c(\omega)) \\
s.t. : \quad & \mathbf{K}\mathbf{u}(\omega) = \mathbf{f}(\omega) \quad (\omega \in \Theta) \\
& \sum_{e=1}^{N} \rho_e v_e \leq \bar{V} \\
& 0 \leq \boldsymbol{\rho} \leq 1
\end{aligned} \tag{11}$$

where $\mu(c(\omega))$ and $\sigma(c(\omega))$ are the expected value and standard deviation of the structural compliance. $\omega$ denotes any realization in a random sampling space $\Theta$. $v_e$ is the volume of element *e* and the total volume of the design is limited to $\bar{V}$. $\beta$ is a weight coefficient used to balance the contribution of the mean and the standard deviation. By increasing the value of $\beta$, the constraint on variability becomes strengthened [46]; when $\beta$ = 0, the objective function corresponds to calculating the minimum expected value of structural compliance. A summary of different kinds of objective functions for robust topology optimization has been presented in [18].

The mean and standard deviation of the structural compliance can be expressed as

$$\mu(c(\omega)) = \int_{\Theta} c(\omega) P d\omega \tag{12}$$

$$\sigma^2(c(\omega)) = \int_{\Theta} [c(\omega) - \mu(c(\omega))]^2 P d\omega \tag{13}$$

where $Pd\omega$ is the probability density function for a continuous random variable in domain $\Theta$.

*3.2. Realization of an imprecise random field load*

As introduced in the previous sections, the traditional exponential function is employed to define covariance. To ensure the accuracy of the truncated K-L expansion[19, 33, 44, 45], a significance check is conducted to choose the most appropriate value of *M*, after which the eigenvalues and eigenfunctions of the corresponding terms are calculated. The details of this procedure are described below.

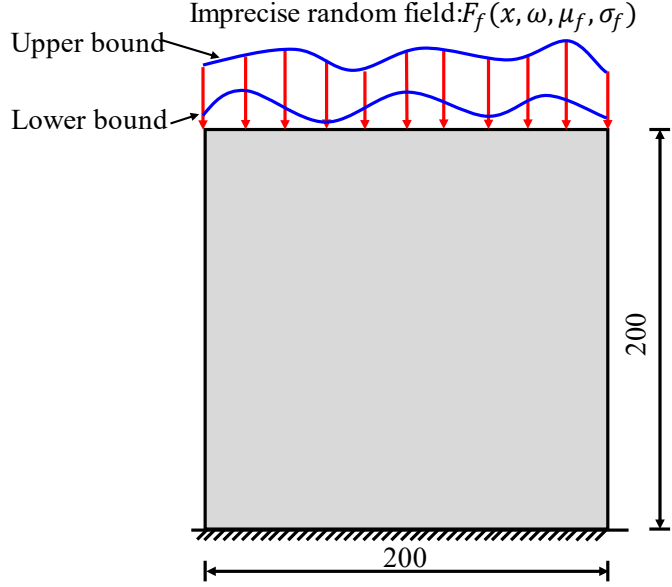

Figure 1 A square plate with imprecise random field loads

Consider a structure subjected to an imprecise random field load, such as Figure 1. The random field load can be represented by the truncated K-L expansion[47, 48] from Eq. (8), as follows.

$$H_f^p(\boldsymbol{x},\omega) = \mu_f^I + \sum_{i=1}^{M} \sqrt{\lambda_i^I} \psi_i(\boldsymbol{x}) \xi_i(\omega) \tag{14}$$

The covariance function for the spatially varying random field is again described by the commonly used exponential function as

$$Cov_f(\boldsymbol{x}_1, \boldsymbol{x}_2) = \left(\sigma_f^I\right)^2 \exp\left(-\frac{|x_1 - x_2|}{L}\right) \tag{15}$$

where $\sigma_f^I$ is the interval standard deviation, $|x_1 - x_2|$ is the distance between any two points and *L* is the given correlation length. The influence of the correlation length is shown in Figure 2. Assuming the domain of the random field is the 1D domain $\Omega = [-a,a] \subset \mathbb{R}^n$, the solution of the eigenfunctions and eigenvalues of the covariance function from the integral equation, is as follows.

$$\left(\sigma_f^I\right)^2 \int_{-a}^{+a} \exp\left(-\frac{|x_1 - x_2|}{L}\right) \psi_i(\pmb{x}_2) d\pmb{x}_2 = \lambda_i^I \psi_i(\pmb{x}_1) \tag{16}$$

where the black bond $\pmb{x}_1$, $\pmb{x}_2$ are the vectors of a series of $x_1$ and $x_2$.

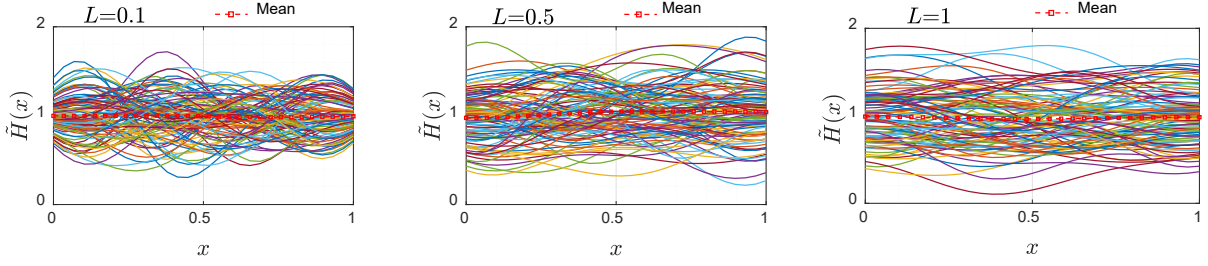

Figure 2 Random field realizations generated by the Gaussian covariance kernel using different correlation lengths

Since the mean and standard deviation of the random field are interval-valued, the eigenvalues and eigenfunctions are also interval-valued and the explicit expressions for the eigenfunctions $\psi_i$ and the eigenvalues $\lambda_i^I$ are given by,

$$\text{For } i \text{ odd, } i \geq 1: \psi_i^*(\pmb{x}) = \frac{\sin\left(\varpi_i^* \pmb{x}\right)}{\sqrt{a - \frac{\sin(2\varpi_i^* a)}{2\varpi_i^*}}}, \lambda_i^{*I} = \frac{2\left(\sigma_f^I\right)^2 L}{1 + \varpi_i^{*2}} \tag{17}$$

$$\text{For } i \text{ even, } i \geq 2: \psi_i(\pmb{x}) = \frac{\cos\left(\varpi_i \pmb{x}\right)}{\sqrt{a + \frac{\sin(2\varpi_i a)}{2\varpi_i}}}, \lambda_i^I = \frac{2\left(\sigma_f^I\right)^2 L}{1 + \varpi_i^2} \tag{18}$$

where $\varpi_i^*$ and $\varpi_i$ are the natural frequencies of the eigenfunctions which can be solved using the following equations, [49].

$$\begin{aligned}\frac{\tan\left(\varpi^* a\right)}{\varpi^* L} + 1 &= 0 \\ \tan\left(\varpi a\right) - \frac{1}{\varpi L} &= 0\end{aligned} \tag{19}$$

From Eq. (14), the imprecise random field load is represented by a linear combination of orthogonal functions. Assuming the loads are applied to the structure one by one and using the linear superposition method we have,

$$\bm{f}_0^I = \mu_f^I, \bm{f}_1^I = \sqrt{\lambda_1^I}\psi_1(\bm{x}), \bm{f}_2^I = \sqrt{\lambda_2^I}\psi_2(\bm{x}),\ldots, \bm{f}_M^I = \sqrt{\lambda_M^I}\psi_M(\bm{x}) \qquad (20)$$

where for each load case, the corresponding displacement fields are $\bm{u}_0^I, \bm{u}_1^I, \bm{u}_2^I, \cdots, \bm{u}_M^I$, which are uniquely determined by $\{\bm{u}^I\} = [\bm{K}]^{-1}\{\bm{f}^I\}$ where $[\bm{K}]$ is the global stiffness matrix. This leads to the realization of the imprecise random field loading based on the sum of all the loads in Eq. (21). Figure 3 illustrates the imprecise random field realization of the load magnitude.

$$\begin{aligned}\bm{f}^I &= \bm{f}_0^I \xi_0(\omega) + \bm{f}_1^I \xi_1(\omega) + \bm{f}_2^I \xi_2(\omega) + \cdots + \bm{f}_M^I \xi_M(\omega) \\ &= \mu_f^I \xi_0(\omega) + \sqrt{\lambda_1^I}\psi_1(\bm{x})\xi_1(\omega) + \sqrt{\lambda_2^I}\psi_2(\bm{x})\xi_2(\omega) + \cdots + \sqrt{\lambda_M^I}\psi_M(\bm{x})\xi_M(\omega)\end{aligned} \qquad (21)$$

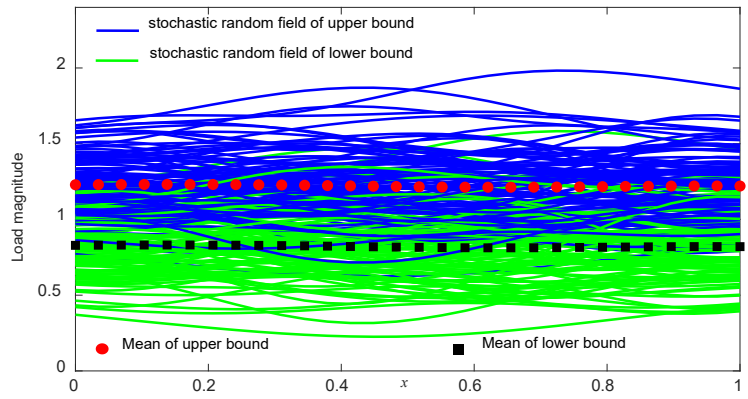

Figure 3 Imprecise random field realizations of load magnitude

*3.3. Calculation of the mean and variance of structural compliance*

For any sample $\omega \in \mathbb{R}^n$, the stochastic compliance of a structure under imprecise field loading $H_f^p(\bm{x}, \omega)$ can be given as,

$$c_1(\omega) = \bm{f}^T(\omega)\bm{u}(\omega) \qquad (22)$$

This is called the direct method. An alternative method is to derive the compliance of the structure and represent the force and displacement vectors as a linear combination of orthogonal functions $\xi_i(\omega)$,

$$\begin{aligned}c_2(\omega) &= \left(\sum_{i=0}^{M} \xi_i(\omega)\bm{f}_i^T\right)\left(\sum_{i=0}^{M} \xi_i(\omega)\bm{u}_i\right) \\ &= \sum_{i=0}^{M}\sum_{j=0}^{M} \xi_i(\omega)\xi_j(\omega)\bm{f}_i^T\bm{u}_j \\ &= \sum_{i=0}^{M}\sum_{j=0}^{M} \xi_i(\omega)\xi_j(\omega)c_{ij}\end{aligned} \qquad (23)$$

where $c_{ij} = \boldsymbol{f}_i^T \boldsymbol{u}_j$ denotes the inner product of vector $\boldsymbol{f}_i$ and vector $\boldsymbol{u}_j$.

The present study uses this second method based on the linear superposition assumption. The efficiency and accuracy of the method is carefully investigated in the following section.

The mean of the compliance from Eq. (23) can be written as,

$$\mu_c = \sum_{i=0}^{M} \sum_{j=0}^{M} \langle \xi_i(\omega) \xi_j(\omega) \rangle c_{ij}$$
$$= \sum_{i=0}^{M} \sum_{j=0}^{M} \delta_{ij} c_{ij} \qquad (24)$$
$$= \sum_{i=0}^{M} c_{ii} = \sum_{i=0}^{M} \boldsymbol{f}_i^T \boldsymbol{u}_i$$

Variance can be derived as, [18, 31, 44]

$$\sigma^2(c) = E(c^2) - \mu^2(c) \qquad (25)$$

where the second term of Eq. (25) is the square of the sum of $c_{ij}$ (not the sum of the squares of $c_{ij}$ as has been reported previously[19]), which is expanded to give,

$$\sigma^2(c) = E(c^2) - \mu^2(c)$$
$$= E\left( \sum_{i,j,k,l=0}^{M} \xi_i(\omega)\xi_j(\omega)\xi_k(\omega)\xi_l(\omega) c_{ij} c_{kl} \right) - \left( \sum_{i=0}^{M} \sum_{j=0}^{M} \delta_{ij} c_{ij} \right)^2$$
$$= \sum_{i,j,k,l=0}^{M} E\left( \xi_i(\omega)\xi_j(\omega)\xi_k(\omega)\xi_l(\omega) \right) c_{ij} c_{kl} - \sum_{i,j,k,l=0}^{M} \delta_{ij} \delta_{kl} c_{ij} c_{kl}$$
$$= \sum_{i,j,k,l=0}^{M} \delta_{ij} c_{ij} \delta_{kl} c_{kl} + \sum_{\substack{i,j,k,l=0 \\ i\neq j}}^{M} \delta_{ik} c_{ij} \delta_{jl} c_{kl} + \sum_{\substack{i,j,k,l=0 \\ i\neq k}}^{M} \delta_{il} c_{ij} \delta_{jk} c_{kl} - \sum_{i,j,k,l=0}^{M} \delta_{ij} \delta_{kl} c_{ij} c_{kl}$$
$$= \sum_{i,j,k,l=0}^{M} \delta_{ij} c_{ij} \delta_{kl} c_{kl} + 2 \sum_{\substack{i,j,k,l=0 \\ i\neq j}}^{M} \delta_{ik} c_{ij} \delta_{jl} c_{kl} - \sum_{i,j,k,l=0}^{M} \delta_{ij} \delta_{kl} c_{ij} c_{kl}$$
$$= 2 \sum_{\substack{i,j,k,l=0 \\ i\neq j}}^{M} \delta_{ik} c_{ij} \delta_{jl} c_{kl}$$
$$= 2 \sum_{\substack{i,j=0 \\ i\neq j}}^{M} c_{ij}^2 \qquad (26)$$

Verified using Isserlis' theorem [50] and taking advantage of the symmetry of the matrices,

$$\begin{aligned}
\sigma^2(c) &= E\left\{[c - E(c)]^2\right\} \\
&= E\left\{\left[\sum_{i=0}^{M}\sum_{j=0}^{M}\xi_i(\omega)\xi_j(\omega)c_{ij} - \sum_{i=0}^{M}c_{ii}\right]^2\right\} = E\left\{\left[\sum_{\substack{i,j=0\\i\neq j}}^{M}\xi_i(\omega)\xi_j(\omega)c_{ij}\right]^2\right\} \\
&= E\left\{\sum_{\substack{i,j=0\\i\neq j}}^{M}\sum_{\substack{k,l=0\\k\neq l}}^{M}\xi_i\xi_j\xi_k\xi_l c_{ij}c_{kl}\right\} = \sum_{\substack{i,j=0\\i\neq j}}^{M}\sum_{\substack{k,l=0\\k\neq l}}^{M}c_{ij}c_{kl}E\left[\xi_i\xi_j\xi_k\xi_l\right] \quad (27)\\
&= \sum_{\substack{i,j=0\\i\neq j}}^{M}\sum_{\substack{k,l=0\\k\neq l}}^{M}c_{ij}c_{kl}\left(\delta_{ik}\delta_{jl} + \delta_{il}\delta_{jk}\right) = 2\sum_{\substack{i,j,k,l=0\\i\neq j}}^{M}\delta_{ik}\delta_{jl}c_{ij}c_{kl} \\
&= 2\sum_{\substack{i,j=0\\i\neq j}}^{M}c_{ij}^2
\end{aligned}$$

where $\xi_{ijkl} = E\left(\xi_i(\omega)\xi_j(\omega)\xi_k(\omega)\xi_l(\omega)\right)$ are the fourth order moments of the random variables. There are two main ways to calculate these moments, one is using the traditional Monte Carlo method [19], the other, more efficient way is as follows. As $\xi_i(\omega)$ are independent random variables with $\xi_i(\omega) \sim N(0, \sigma_i^2)$ for $i \in \mathbb{R}^n$ [51], we can have,

$$E\left(\xi_i\xi_j\xi_k\xi_l\right) = \begin{cases} 3\sigma_i^4 & \text{if} \quad i = j = k = l \\ \sigma_i^2\sigma_j^2 & \text{if} \quad i = k, j = l \quad \text{or} \quad i = l, j = k \quad \text{and} \quad i \neq j \\ \sigma_i^2\sigma_k^2 & \text{if} \quad i = j, k = l \quad \text{and} \quad i \neq k \\ 0 & \text{other cases} \end{cases} \quad (28)$$

## 4. Sensitivity analysis with imprecise random field loads

The mean and standard deviation of structural compliance are interval uncertainties, leading to an imprecise objective function expressed as

$$\begin{bmatrix} \underline{J} & \overline{J} \end{bmatrix} = \begin{bmatrix} \underline{\mu} & \overline{\mu} \end{bmatrix} + \beta\begin{bmatrix} \underline{\sigma} & \overline{\sigma} \end{bmatrix} \quad (29)$$

Authors such as Wu et al. [27] dealt with such uncertainty in optimization by considering the worst case optimization, using the upper bound of the objective function as the new objective function. However, the numerical examples presented in their work show that the optimization convergence was often unstable. In this study, we present a novel method for a sensitivity analysis with a bounded compliance

$$\frac{\partial J}{\partial \rho} = w_1 \frac{\partial \overline{J}}{\partial \rho} + w_2 \frac{\partial \underline{J}}{\partial \rho} \quad (30)$$

where $w_1$, $w_2$ are weight coefficients used to balance the contribution of the upper and lower bounds. When $w_1 = 0, w_2 = 1$, minimum compliance is considered, while when $w_1 = 1, w_2 = 0$, compliance is maximum.

*4.1. Traditional sensitivity analysis*

The deterministic structural compliance can be expressed as,

$$C = \mathbf{F}^T\mathbf{U} = \mathbf{U}^T\mathbf{K}\mathbf{U} \tag{31}$$

where $\mathbf{F}, \mathbf{U}$ are the design-independent load vector and global displacement vector respectively and $\mathbf{K}$ is the global stiffness matrix. Utilizing the well-known adjoint method, the sensitivity of the structural compliance with respect to the design variable $\rho_e$ can be derived as,

$$\begin{aligned}\frac{\partial C}{\partial \rho_e} &= \frac{\partial}{\partial \rho_e}\left[\mathbf{U}^T\mathbf{K}\mathbf{U} + \chi(\mathbf{K}\mathbf{U} - \mathbf{F})\right] \\ &= 2\mathbf{U}^T\mathbf{K}\frac{\partial \mathbf{U}}{\partial \rho_e} + \mathbf{U}^T\frac{\partial \mathbf{K}}{\partial \rho_e}\mathbf{U} + \chi\left(\frac{\partial \mathbf{K}}{\partial \rho_e}\mathbf{U} - \mathbf{K}\frac{\partial \mathbf{U}}{\partial \rho_e}\right)\end{aligned} \tag{32}$$

As $\chi$ can be any adjoint vector, we assume $\chi = -2\mathbf{U}^T$. Eq. (32) can then be reformulated as,

$$\frac{\partial C}{\partial \rho_e} = -\mathbf{U}^T\frac{\partial \mathbf{K}}{\partial \rho_e}\mathbf{U} \tag{33}$$

For nondeterministic problems, a gradient-based method is applied to the objective function in Eq. (11). Based on the chain rule, sensitivities with respect to physical variables can be given by,

$$\begin{aligned}\frac{\partial J}{\partial \rho_e} &= \frac{\partial \mu_c}{\partial \rho_e} + \beta\frac{\partial \sigma(c)}{\partial \rho_e} = \frac{\partial \mu_c}{\partial \rho_e} + \frac{\beta}{2\sigma(c)}\frac{\partial \sigma^2(c)}{\partial \rho_e} \\ &= \sum_{i,j=0}^{M}\delta_{ij}\frac{\partial c_{ij}}{\partial \rho_e} + \frac{\beta}{2\sigma(c)}\left[\sum_{i,j,k,l=0}^{M}\left(\xi_{ijkl} - \delta_{ij}\delta_{kl}\right)\left(\frac{\partial c_{ij}}{\partial \rho_e}c_{kl} + \frac{\partial c_{kl}}{\partial \rho_e}c_{ij}\right)\right] \\ &= \sum_{i,j=0}^{M}\delta_{ij}\frac{\partial c_{ij}}{\partial \rho_e} + \frac{\beta}{\sigma(c)}\left[\sum_{i,j,k,l=0}^{M}\left(\xi_{ijkl} - \delta_{ij}\delta_{kl}\right)\frac{\partial c_{ij}}{\partial \rho_e}c_{kl}\right] \\ &= \sum_{i,j=0}^{M}\left[\delta_{ij} + \frac{\beta}{\sigma(c)}\sum_{k,l=0}^{M}\left(\xi_{ijkl} - \delta_{ij}\delta_{kl}\right)c_{kl}\right]\frac{\partial c_{ij}}{\partial \rho_e}\end{aligned} \tag{34}$$

Defining,

$$w_{ij} = \delta_{ij} + \frac{\beta}{\sigma(c)}\sum_{k,l=0}^{M}\left(\xi_{ijkl} - \delta_{ij}\delta_{kl}\right)c_{kl} \tag{35}$$

and substituting Eq. (29) into Eq. (30) and using the discrete linear elasticity, this can be revised to,

$$\frac{\partial J}{\partial \rho_e} = \sum_{i,j=0}^{M} w_{ij} \frac{\partial c_{ij}}{\partial \rho_e}$$
$$= -\sum_{i,j=0}^{M} w_{ij} \mathbf{U}_i^{\mathrm{T}} \frac{\partial \mathbf{K}}{\partial \rho_e} \mathbf{U}_j \qquad (36)$$
$$= -p\rho_e^{p-1}(E_0 - E_{\min}) \sum_{i,j=0}^{M} w_{ij} \mathbf{u}_{ie}^{\mathrm{T}} \mathbf{k}_e \mathbf{u}_{je}$$

where $\mathbf{u}_{ie}$ is the nodal displacement of the $e$-th element and $\mathbf{k}_e$ is the stiffness matrix of element $e$ with solid material. To solve the above equation, $(M+1)^2$ load cases need to be calculated for the sensitivity analysis.

*4.2. Interval sensitivity analysis with imprecise random field*

The derivatives of the objective function for both the upper and lower bounds can be expressed as,

$$\frac{\partial \overline{J}}{\partial \rho_e} = -p\rho_e^{p-1}(E_0 - E_{\min}) \sum_{i,j=0}^{M} \overline{w}_{ij} \overline{\mathbf{u}}_{ie}^{\mathrm{T}} \mathbf{k}_e \overline{\mathbf{u}}_{je}$$
$$\frac{\partial \underline{J}}{\partial \rho_e} = -p\rho_e^{p-1}(E_0 - E_{\min}) \sum_{i,j=0}^{M} \underline{w}_{ij} \underline{\mathbf{u}}_{ie}^{\mathrm{T}} \mathbf{k}_e \underline{\mathbf{u}}_{je} \qquad (37)$$

where $\overline{\mathbf{u}}_{ie}$, $\underline{\mathbf{u}}_{ie}$ are the upper and lower nodal displacements of the $i$-th element, respectively. As can be seen, for each of the load cases, the corresponding displacement fields need to be solved, in addition to the other parameters involved. Moreover, for each iteration of the sensitivity analysis, such a step needs to be repeated. For example, if it takes $N$ steps to meet the convergence criteria, then a total $(M+1)^{2N}$ steps will be needed, in addition to searching the upper and lower bounds of the indispensable parameters.

In order to improve the computational efficiency, a number of techniques are introduced as follows. The first is the Orthogonal Similarity Transformation (OST) method which is used to reduce the computational cost of calculating $w_{ij}$. Observation of Eq. (35), reveals that the elements of $w_{ij}$ form a real symmetric matrix $\mathbf{W}_{(M+1)\times(M+1)}$ and that there exists an orthogonal matrix $\mathbf{T}$ which makes $\mathbf{W}$ into a diagonal matrix,

$$\mathbf{T}^T \mathbf{W} \mathbf{T} = \mathbf{T}^{-1} \mathbf{W} \mathbf{T} = diag(\lambda_0, \lambda_1, ..., \lambda_M) \qquad (38)$$

where $T$ is an orthogonal matrix with column vectors $q_i$ and $\lambda_1, \lambda_2, ..., \lambda_n$ are the eigenvalues of $W$. Introducing the following matrix to simplify the derivation process,

$$\begin{aligned} \boldsymbol{f}_t &= \left[\{\boldsymbol{f}_0\}, \{\boldsymbol{f}_1\}, ..., \{\boldsymbol{f}_M\}\right] & \boldsymbol{F}_t &= \left[\{\boldsymbol{F}_0\}, \{\boldsymbol{F}_1\}, ..., \{\boldsymbol{F}_M\}\right] = \boldsymbol{f}_t \boldsymbol{T}_{(M+1)\times(M+1)} \\ \boldsymbol{u}_t &= \left[\{\boldsymbol{u}_0\}, \{\boldsymbol{u}_1\}, ..., \{\boldsymbol{u}_M\}\right] & \boldsymbol{U}_t &= \left[\{\boldsymbol{U}_0\}, \{\boldsymbol{U}_1\}, ..., \{\boldsymbol{U}_M\}\right] = \boldsymbol{u}_t \boldsymbol{T}_{(M+1)\times(M+1)} \end{aligned} \quad (39)$$

where $t = 0, 1, ..., M$ and $\boldsymbol{f}_t, \boldsymbol{u}_t$ are the $1 \times (M+1)$ matrices of the load cases and the corresponding displacements for each degree of freedom. Therefore, by substituting Eq. (38) and Eq. (39) into Eq. (34), one can obtain,

$$\begin{aligned} \frac{\partial J}{\partial \rho_e} &= \frac{\partial}{\partial \rho_e} \sum_{i,j=0}^{M} \left(w_{ij} c_{ij}\right) = \frac{\partial}{\partial \rho_e} tr\left(\boldsymbol{W} \boldsymbol{u}_t^T \boldsymbol{f}_t\right) = \frac{\partial}{\partial \rho_e} tr\left(\boldsymbol{f}_t \boldsymbol{W} \boldsymbol{u}_t^T\right) \\ &= \frac{\partial}{\partial \rho_e} tr\left(\boldsymbol{f}_t \boldsymbol{T} diag(\lambda_0, \lambda_1, ..., \lambda_M) \boldsymbol{T}^T \boldsymbol{u}_t^T\right) \\ &= \frac{\partial}{\partial \rho_e} tr\left(\boldsymbol{F}_t diag(\lambda_0, \lambda_1, ..., \lambda_M) \boldsymbol{U}_t^T\right) \\ &= \sum_{t=0}^{M} \lambda_t \frac{\partial}{\partial \rho_e} tr\left(\boldsymbol{F}_t \boldsymbol{U}_t^T\right) \\ &= \sum_{i=0}^{M} \lambda_i \frac{\partial \left(\boldsymbol{F}_i^T \boldsymbol{U}_i\right)}{\partial \rho_e} \\ &= -p \rho_e^{p-1} (E_0 - E_{\min}) \sum_{i=0}^{M} \lambda_i \boldsymbol{u}_{ie}^T \boldsymbol{k}_e \boldsymbol{u}_{ie} \end{aligned} \quad (40)$$

where $tr(\bullet)$ denotes the sum of the diagonal elements, and the new sensitivity analysis with a bounded compliance can be further derived as,

$$\frac{\partial \tilde{J}}{\partial \rho} = -w_1 p \rho_e^{p-1} (E_0 - E_{\min}) \sum_{i=0}^{M} \bar{\lambda}_i \bar{\boldsymbol{u}}_{ie}^T \boldsymbol{k}_e \bar{\boldsymbol{u}}_{ie} - w_2 p \rho_e^{p-1} (E_0 - E_{\min}) \sum_{i=0}^{M} \underline{\lambda}_i \underline{\boldsymbol{u}}_{ie}^T \boldsymbol{k}_e \underline{\boldsymbol{u}}_{ie} \quad (41)$$

### 4.3. Improved Heaviside filtering

To avoid numerical instability and obtain black-and-white solutions, the present paper adopts a volume preserving nonlinear density filter [52] which combines the Heaviside filter with a given volume, resulting in the volumes being the same before and after filtering. This formulation not only preserves the advantages of the filter but also improves the convergence rate.

Using linear density filtering, the filtered density $\bar{\rho}_e$ for the nth element can be expressed as,

$$\bar{\rho}_e = \frac{\sum_{i \in N_e} w(\mathbf{x}_i) v_i \rho_i}{\sum_{i \in N_e} w(\mathbf{x}_i) v_i} \tag{42}$$

where $\rho_i$ are the original design variables and $v_i$ denotes the volume of the $i$-th element. $N_e$ are the elements around e whose centres are located within $R$, and the weighting coefficient can be defined as,

$$w(\mathbf{x}_i) = R - \|\mathbf{x}_i - \mathbf{x}_e\| \tag{43}$$

The filtered density $\bar{\rho}_e$ is further filtered by the volume preserving nonlinear density filter [52] and the physical density is,

$$\tilde{\rho}_e = \begin{cases} \eta \left[ e^{-\alpha(1-\bar{\rho}_e/\eta)} - (1-\bar{\rho}_e/\eta) e^{-\alpha} \right] & 0 \leq \bar{\rho}_e \leq \eta \\ (1-\eta)\left[ 1 - e^{-\alpha(\bar{\rho}_e/\eta)/(1-\eta)} + (\bar{\rho}_e - \eta) e^{-\alpha}/(1-\eta) \right] + \eta & \eta \leq \bar{\rho}_e \leq 1 \end{cases} \tag{44}$$

where $\alpha$ is the smooth parameter in the Heaviside function and $\eta$ is the threshold value which controls the volume before and after filtering and can be determined using the bisection method according to the volume preserving condition. The sensitivities of the general function $f$ with respect to its original density $\rho_e$ can be obtained by the chain rule

$$\frac{\partial f}{\partial \rho_e} = \sum_{i \in N_e} \frac{\partial f}{\partial \tilde{\rho}_i} \frac{\partial \tilde{\rho}_i}{\partial \bar{\rho}_i} \frac{\partial \bar{\rho}_i}{\partial \rho_e} \tag{45}$$

The well-known globally convergent method of moving asymptotes (GCMMA) [53] is utilized to solve the robust topology optimization with imprecise random field loads. This is particularly suitable for large scale optimization problems, decomposing them into several single variable problems.

## 5. Solution procedure for statistical responses with imprecise random field loads

*5.1. Framework for robust topology optimization with imprecise random field loads*

The solution algorithm for robust topology optimization with imprecise random field loads is summarized by the flowchart shown in Figure 4. The main steps are as follows.

*Step 1.* Initialization: The uncertain variables and the mean and standard deviation of an external force are derived from $n_s$ samples. Three different percentages of confidence interval (CI), for example 90%, 95% and 99% are then considered to compute interval bounds for the $\Xi^p \sim \left( \left[ \underline{\mu}_f, \bar{\mu}_f \right], \left[ \underline{\sigma}_f, \bar{\sigma}_f \right] \right)$ can be formulated as

$$\Xi^p = \left\{ F_f(\boldsymbol{x}, \omega; \mu_f, \sigma_f) \middle| \forall \boldsymbol{x} \in \mathbb{R}^n, \mu_f \in \left[ \underline{\mu}_f, \bar{\mu}_f \right], \sigma_f \in \left[ \underline{\sigma}_f, \bar{\sigma}_f \right] \right\} \quad (46)$$

*Step 2*. Discretization of the random field loading: Here, the traditional exponential function is employed to define the covariance function. To ensure the accuracy of the truncated K-L expansion, a significance check is conducted to choose the most appropriate value of *M* using Eq. (47). Then the eigenvalues and eigenfunctions of the corresponding terms are calculated, as described in section 3.2.

$$s = \sum_{i=1}^{M} \lambda_i \bigg/ \sum_{i=1}^{N} \lambda_i \quad (47)$$

*Step 3*. Boundary search: Using an FE analysis the displacement of the corresponding realization of the imprecise random field load is calculated. The mean and standard deviation of the compliance are obtained based on each sampling using Eqs. (24) and (26); then the bounds and output $\left[ \underline{u} \quad \bar{u} \right]$, $\left[ \underline{J} \quad \bar{J} \right]$, $\left[ \underline{\mu} \quad \bar{\mu} \right]$, $\left[ \underline{\sigma} \quad \bar{\sigma} \right]$, $\left[ \underline{c}_{ij} \quad \bar{c}_{ij} \right]$, $\left[ \underline{\lambda} \quad \bar{\lambda} \right]$ are searched; Further detail on this step is discussed in the next subsection.

*Step 4*. Interval sensitivity analysis: By substituting the bounded parameters from *Step* 3, the upper and lower bounds of the objective function are obtained using the OST method after which an interval sensitivity analysis is performed by Eq. (41).

*Step 5*. Filtering of the sensitivities based on section 4.3.

*Step 6*. Updating the density, checking the constraint and convergence: If both the volume constraint and convergence are satisfied, stop the loop and output the final robust design; if not, go back to *Step* 3.

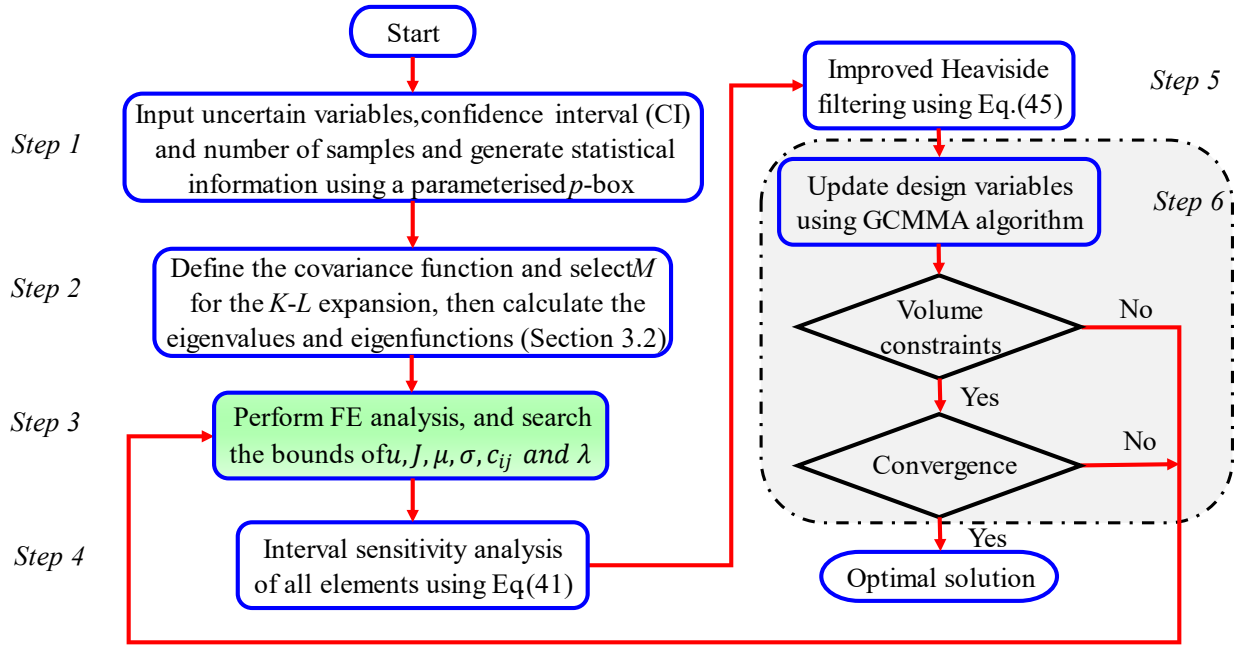

Figure 4 Flowchart of the solution procedure of topology optimization with imprecise random field

## 5.2. Boundary search of important parameters

Three different methods for searching the boundaries of the important intermediate parameters namely QMCS, LHNPSO and CA are detailed in this section.

- Quasi-Monte Carlo Simulation (QMCS)

The most straightforward approach used to obtain statistical information on stochastic systems is the Monte Carlo sampling strategy. By generating a large group of random field realizations of uncertain variables and performing a deterministic analysis at each of these, a user can get reliable results by using sufficient numerical computations. This method however, presents a challenge for a large-scale analysis. Quasi-Monte Carlo simulation possesses shorter computational times and higher accuracy by utilizing quasi-random sequences rather than random sampling. QMCS is employed here as the reference method. The present QMCS approach adopts a low-discrepancy Sobol sequence by skipping the first 1000 values and retaining every 101st point after that to generate the uncertain variables. For each iteration of the topology optimization, a QMCS is implemented in *Step* 3 of Figure 4 to find the bounded parameters.

- LHNPSO algorithm

The PSO method is a population based stochastic optimization technique used to find the optimal solutions and paths within a given interval. In this problem *Step* 3 can be transformed into an optimization problem as follows:

$$\begin{aligned}&\min_{\mu_f,\sigma_f}: \quad P_{\min}(\omega) \quad (\omega \in \Theta)\\ &\max_{\mu_f,\sigma_f}: \quad P_{\max}(\omega) \\ &s.t.: \begin{cases} \underline{\mu}_f \leq \mu_f(\omega) \leq \overline{\mu}_f \\ \underline{\sigma}_f \leq \sigma_f(\omega) \leq \overline{\sigma}_f \end{cases}\end{aligned} \qquad (48)$$

where $P$ is the bounded parameter in Step.3, i.e. $[\underline{u}\ \overline{u}]$, $[\underline{J}\ \overline{J}]$, $[\underline{\mu}\ \overline{\mu}]$, $[\underline{\sigma}\ \overline{\sigma}]$, $[\underline{c}_{ij}\ \overline{c}_{ij}]$ or $[\underline{\lambda}\ \overline{\lambda}]$ based on a collection of random field realizations and the imprecise probability load distribution is $\Xi^p \sim ([\underline{\mu}_f, \overline{\mu}_f], [\underline{\sigma}_f, \overline{\sigma}_f])$.

We adopt the low-discrepancy sequence initialized high-order nonlinear particle swarm optimization (LHNPSO) method by Yang et al. [55]. Their results indicated that their PSO method could converge faster and offered the more accurate solutions than the conventional method. The velocity $Vel_v^{t+1}$ and position vectors $Pos_v^{t+1}$ of sample $v$ at time $t+1$ can be derived by the following equations, respectively.

$$\begin{aligned} Vel_v^{t+1} &= \varpi Vel_v^t + c_1 r_1 \left(Pos_v^{pbest} - Pos_v^t\right) + c_2 r_2 \left(Pos^{gbest} - Pos_v^t\right) \\ Pos_v^{t+1} &= Pos_v^t + Vel_v^{t+1} \end{aligned} \qquad (49)$$

where $Pos_v^{pbest}$ and $Pos^{gbest}$ are the best local and global best solutions. Parameters $c_1, c_2$ are equal to 2. $r_1, r_2$ are two random numbers. Initial weight $\varpi$ is selected based on reference [55]. By using the LHNPSO algorithm, the computational time required to search the bounded parameters in *Step* 3 is drastically reduced compared to the other two methods.

- Combinatorial approach

The combinatorial approach was first presented by Rao and Berke [35] for interval estimates and involved using a combination of the extreme values. The success of this method is due to the monotonicity of the properties of a stochastic system within the range of concern. A *k*-dimensional interval vector [**X**], can be expressed as

$$[\mathbf{X}] = [\underline{\mathbf{X}}, \overline{\mathbf{X}}] = \{x_i : \underline{x}_i \leq x_i \leq \overline{x}_i, i=1,2,...,k\}, [\mathbf{X}] \in \Re^k \qquad (50)$$

where the notations $\underline{\mathbf{X}} = [\underline{x}_1,...,\underline{x}_i...,\underline{x}_k]^{\mathrm{T}}$ and $\overline{\mathbf{X}} = [\overline{x}_1,...,\overline{x}_i...,\overline{x}_k]^{\mathrm{T}}$ represent the lower and upper bound vectors of a set of interval variables, respectively. Therefore, without loss of generality, for an arbitrary function $f$ with $k$-dimensional interval variables $x_i$, all possible values of $f$ can be described as

$$f_r = f\left(x_1^{(1,2)},...,x_i^{(1,2)},...,x_k^{(1,2)}\right); r = 1, 2, ..., 2^k \qquad (51)$$

where $x_1^{(1,2)},...,x_i^{(1,2)},...,x_k^{(1,2)}$ denote the upper and lower bounds of $x_i$, respectively. The system responses can then be obtained as

$$\begin{aligned}\underline{f_r} &= \min\left\{f : f\left(x_1^{(1,2)},...,x_i^{(1,2)},...,x_k^{(1,2)}\right), x_i \in [\mathbf{X}]\right\}, i = 1, 2, ..., k \\ \overline{f_r} &= \max\left\{f : f\left(x_1^{(1,2)},...,x_i^{(1,2)},...,x_k^{(1,2)}\right), x_i \in [\mathbf{X}]\right\}, i = 1, 2, ..., k\end{aligned} \qquad (52)$$

Verification of the monotonicity of the properties in this study will be presented in the following section. The implementation of CA for interval estimates can significantly improve the computational efficiency.

## 6. Numerical examples

In this section, the validity, accuracy, as well as the applicability of the proposed method are demonstrated based on three benchmark examples by comparing the upper and lower bounds of the mean and standard deviation of compliance and the topological layouts. The initialization of the uncertain variables and the mean and standard deviation of the external force are derived from $n_s$ samplings with a system model then generated using the parameterised $p$-box introduced in Eq. (20). Spatial fluctuations of the external loads are modelled as imprecise random fields and a truncated $K$-$L$ expansion is employed to obtain the realization of the random field loading. A significance check is conducted to choose an appropriate value of $M$ using Eq. (47) to capture 90% of the spatial fluctuations of the domain. First, the linear superposition method is compared with the direct method. To implement the CA, the monotonic property of the stochastic system is carefully examined for the interval variables. Then, the accuracy of the proposed RTO algorithm is examined by comparison with the LHNPSO approach and the QMCS method, respectively. Finally, two examples based on periodic structures are presented in detail to show the effectiveness and feasibility of the proposed method. The results demonstrate the method's ability to obtain more robust optimal solutions. Comparison with other methods also reveals that the algorithm is faster, more efficient and more flexible. All of the presented numerical examples were conducted utilizing a workstation with an Intel ®Core™ i9-9900K CPU and 128 GB of RAM.

6.1. *Numerical experiment 1: Carrier plate design*

The carrier plate design problem is investigated, where a plate is fixed along its bottom edge with an imprecise random field load applied to the top edge, as shown in Figure 1. The dimensions of the structure are 200×200, and it is meshed using 200 by 200 linear quadrilateral elements. The Young's modulus of the solid and void materials are 1000 and $10^{-9}$, respectively. The Poisson's ratio is 0.3. The improved Heaviside filtering uses a filter radius of 3 and the volume fraction is set to 0.3. The mean and standard deviation of the load magnitude are -1 and 1.5, respectively. Three different percentages of confidence interval, i.e. 90%, 95% and 99% are chosen to represent the interval mean and standard deviation as indicated in subsection 5.1. The variation of the load magnitude is modelled as a Gaussian random field and an exponential function is employed to define the covariance function, Eq. (15). The correlation length of the distributed loads in this experiment is 10. $M$ for the truncated K-L expansion is selected after the significance check as 14, based on consideration of 90% of the energy field.

6.1.1 *Verification of the linear superposition method*

As discussed in section 3.3, two methods are used to obtain the stochastic compliance due to random inputs for comparison. The first one, called the direct method, generates random variables and each of the compliances is calculated. The second method represents the force and displacement vectors of the compliance as a linear combination of orthogonal functions with limited terms. The efficiency and accuracy of the second method are investigated here:

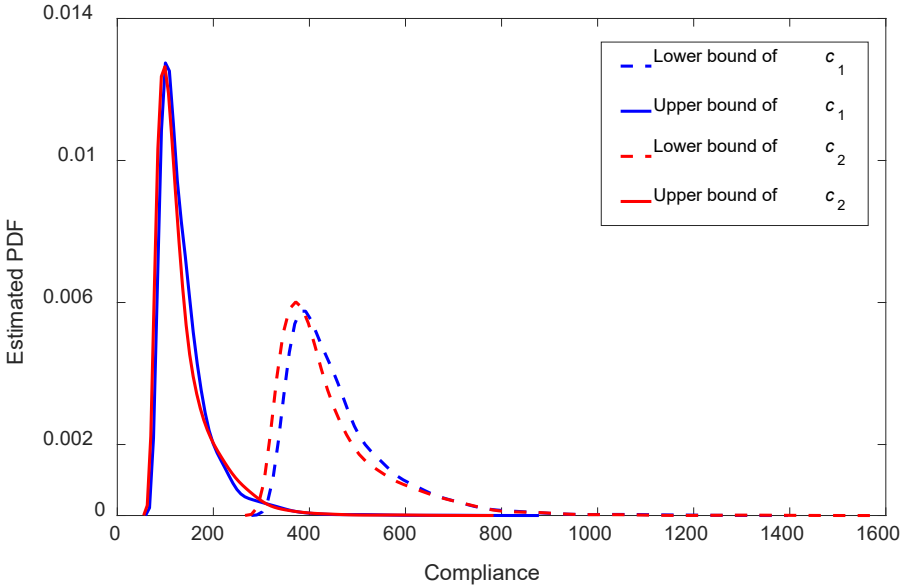

Figure 5 (a) Comparison of the upper and lower bounds of $c_1$ and $c_2$ for the PDF

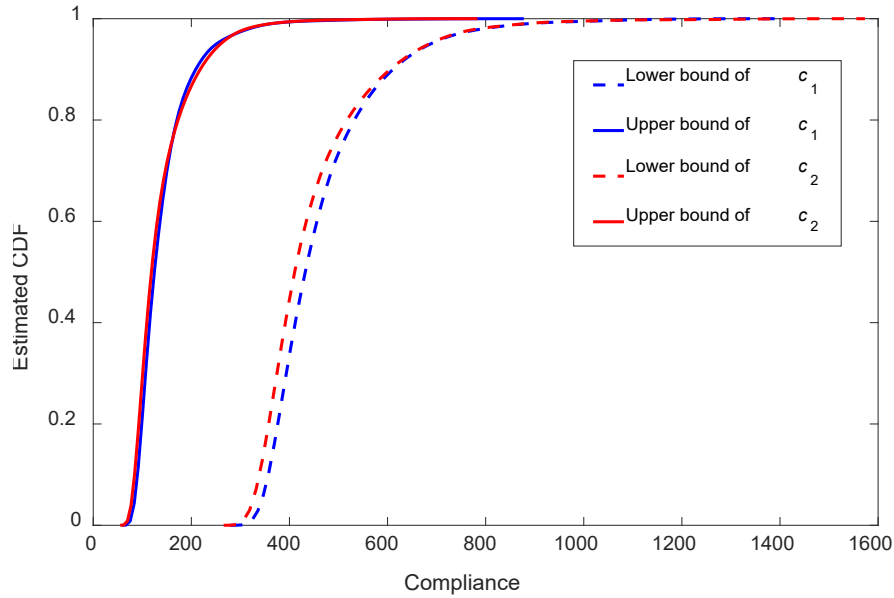

Figure 5 (b) Comparison of the upper and lower bounds of $c_1$ and $c_2$ for the CDF

The probability distribution functions (PDFs) and cumulative distribution functions (CDFs) of the compliances are compared in Figure 5. For the direct method, 1000 samples were chosen to show the efficiency and accuracy of the linear superposition method. As can be seen, the PDFs and CDFs given by the linear superposition method match very well with those obtained by the direct method with the difference between them being around 2%. For the direct method, compliance is obtained after 10 minutes; while for the linear superposition method, 1 minute is needed. Thus, the efficiency and accuracy of the linear superposition method are clearly demonstrated.

6.1.2 *Verification of the monotonicity assumption*

To implement the combinatorial approach in this study, the assumption of the monotonic property of stochastic interval analysis is verified in this subsection. Whilst this assumption has already been accepted by many scholars for linear static structural analysis in engineering applications[35, 42, 56], it is difficult to obtain an explicit mathematical proof directly. In this study, a numerical examination of the monotonic property is adopted. For an arbitrary interval variable, structural responses are thoroughly calculated across the whole range with other interval variables taking their mean values. Then the signs of the derivatives of the target solutions are employed to check the monotonic property. This step is repeated until all of the interval variables are validated. The objective function is influenced by two interval variables, the mean $\mu_f \in \left[\underline{\mu}_f \quad \overline{\mu}_f\right]$ and the standard deviation $\sigma_f \in \left[\underline{\sigma}_f \quad \overline{\sigma}_f\right]$, and the corresponding responses are the mean $\mu(c)$ and the standard deviation $\sigma(c)$ of the compliance. The derivative of a function can be expressed numerically as,

$$\frac{\partial f}{\partial x} = \frac{f_a - f_b}{x_a - x_b} \tag{53}$$

where $x_a$ and $x_b$ are the upper and lower bounds of the interval variables, respectively. $f_a$ and $f_b$ denote the corresponding responses at $x_a$ and $x_b$. With the increase of the sampling points from the interval range, the derivative of the function or slope in Figure 6 becomes accurate. The following four cases are studied.

$$\mu(c) \ \ VS \ \ \mu_f; \quad \sigma(c) \ \ VS \ \ \mu_f$$
$$\mu(c) \ \ VS \ \ \sigma_f; \quad \sigma(c) \ \ VS \ \ \sigma_f$$

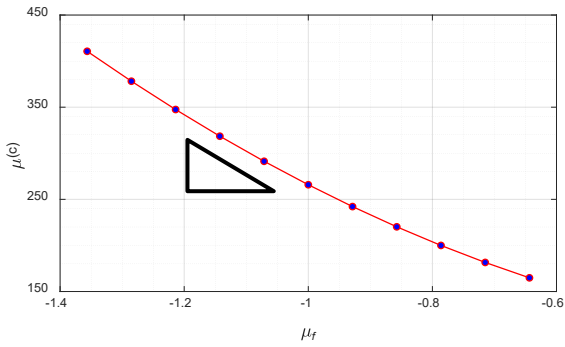 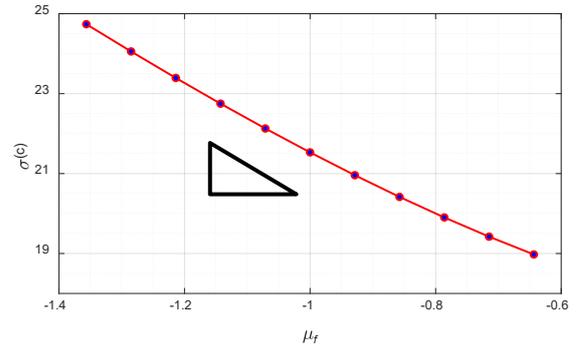

(a) The effect of changing $\mu_f$ on $\mu(c)$      (b) The effect of changing $\mu_f$ on $\sigma(c)$

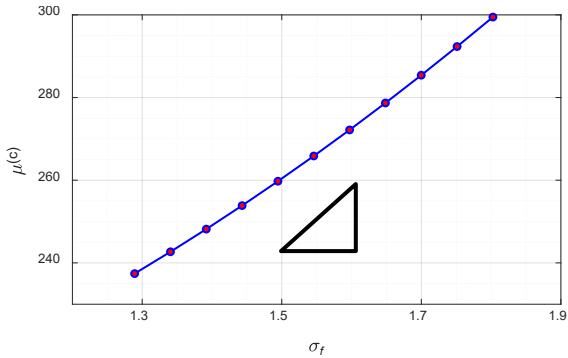 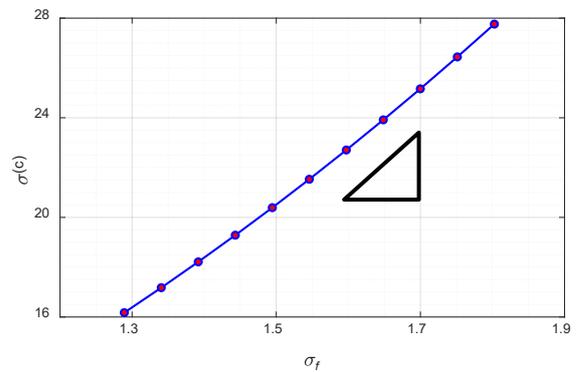

(c) The effect of changing $\sigma_f$ on $\mu(c)$      (d) The effect of changing $\sigma_f$ on $\sigma(c)$

Figure 6 The monotonicity of parameters $\mu(c)$ and $\sigma(c)$

Figure 6 shows the monotonic properties of parameters $\mu(c)$ and $\sigma(c)$. $\sigma_f$ is taken at its midpoint according to the given ranges when examining the influence of $\mu_f$ on $\mu(c)$ and $\sigma(c)$, respectively. A series of $\mu_f$ are carefully observed. A similar procedure is employed for the influence of $\sigma_f$ on $\mu(c)$ and $\sigma(c)$, respectively. The results show that the derivatives or the slopes of Figure 6(c) and (d) are unchanged within the range of $\mu_f$ and $\sigma_f$. Hence the proposed CA can be applied to study the imprecise random field, the mean $\mu_f \in \left[\underline{\mu}_f \ \ \overline{\mu}_f\right]$ and the standard deviation $\sigma_f \in \left[\underline{\sigma}_f \ \ \overline{\sigma}_f\right]$.

### 6.1.3 *Comparison of QMCS, PSO, CA and DTO*

As stated in Section 5.2, three different methods are proposed to search the boundaries of important parameters, including $[\underline{u}\ \ \bar{u}]$, $[\underline{J}\ \ \bar{J}]$, $[\underline{\mu}\ \ \bar{\mu}]$, $[\underline{\sigma}\ \ \bar{\sigma}]$, $[\underline{c}_{ij}\ \ \bar{c}_{ij}]$ and $[\underline{\lambda}\ \ \bar{\lambda}]$. The results obtained using PSO and CA are verified by comparison with those from QMCS, which utilizes a low-discrepancy Sobol sequence b to generate the uncertain variables. Using this method, it can take hundreds of iterations for topology optimization to reach an optimal design. For each step, a boundary search needs to be implemented. For a fair and meaningful comparison, therefore, the first step of the topology optimization process is investigated comparing the bounded parameters and computational times of the different methods, as shown in Table 1. Here, 10,000 samples from QMCS are considered. Statistical information on the responses and computational effort are reported. The results of both the PSO and CA match very well with the QMCS method, although the upper bounds obtained by the QMCS are slightly lower than those of the PSO or CA and the lower bounds of the QMCS are slightly larger. Thus, the accuracy of PSO and CA are demonstrated and the results from these two methods completely encompass the results of the QMCS. By increasing the number of sampling points in the QMCS, the final results are expected to approach those of the PSO or CA methods. With regard to time, the proposed CA method takes 16 seconds for one loop, while the QMCS needs approximately 13 hours and the PSO method takes over 2 minutes. Thus, CA exhibits a high level of accuracy and efficiency.

Table 1 The comparison of bounded parameters and computational times of different methods at the first step

| Methods | *J(c)* | | *μ(c)* | | *σ(c)* | | Computational time (s) |
|---|---|---|---|---|---|---|---|
| | Upper | Lower | Upper | Lower | Upper | Lower | |
| QMCS-10000 | 99.666 | 29.828 | 90.624 | 25.710 | 9.042 | 4.118 | 46223.455 |
| LHNPSO | 100.079 | 29.635 | 91.017 | 25.528 | 9.062 | 4.108 | 144.565 |
| CA | 100.079 | 29.635 | 91.017 | 25.528 | 9.062 | 4.108 | 16.509 |

The results of a deterministic topology optimization (DTO), LHNPSO and CA are compared in Figure 7, Figure 8 and Table 2. Here $w_1 = 1, w_2 = 0$, reducing the CA method in this case to the *minimax* optimization problem [27]. A 90% confidence interval is considered, and *M* is taken as 14 for the truncated the *K-L* expansion, with the imprecise probability load distribution modelled as a Gaussian random field, that is $\Xi^p \sim (\mu \in [-1.356, -0.644], \sigma \in [1.289, 1.803])$. The robust final designs of the DTO, LHNPSO and CA are shown in Figure 7. The robust designs of the LHNPSO and CA are almost the same, with an enhanced central section appearing in the robust designs but not in the DTO design. A comparison of the bounded PDFs and CDFs of the compliances from DTO, LHNPSO and CA, is made in Figure 8. Compared with pure randomness resulting in a singular PDF and CDF, the imprecise randomness introduced in this study incorporates random boundaries encompassing all possible cases across the upper and lower bounds of the compliances.

Under the same level of uncertainty, the objective function values, means and standard deviations of compliances from both LHNPSO and CA are nearly identical, as can be seen in Table 2. The values from the DTO, such as the objective function values and the mean and standard deviation of the compliance however, are greater than those of LHNPSO and CA. For example, the upper bound of $J(c)$ from DTO is 551.99, while by using the proposed method, the corresponding value is 475.23. Although both the LHNPSO and the CA can provide sound and reliable solutions, the computational cost of LHNPSO is much higher than that for the CA, 26.5 hrs versus 1.4 hrs. The superior properties of CA are clearly evident from these examples.

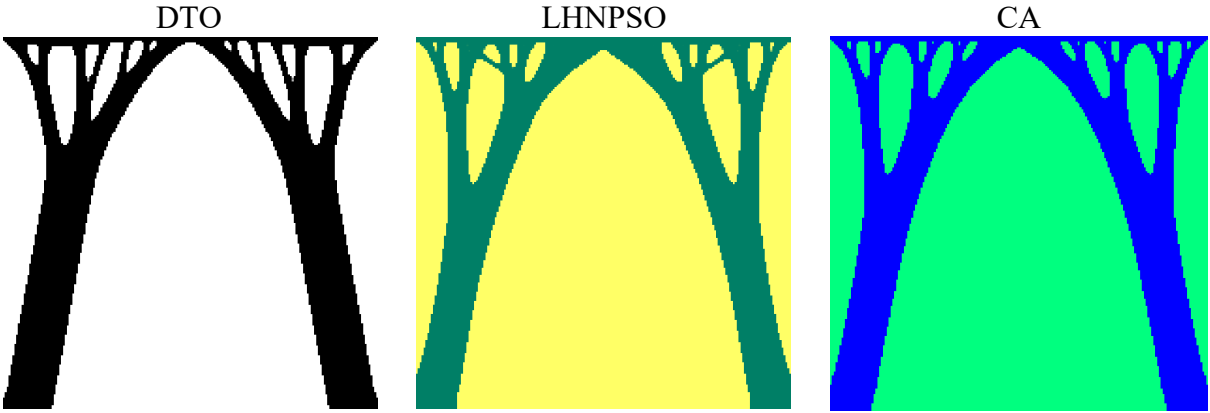

Figure 7 The final design of DTO, LHNPSO and CA

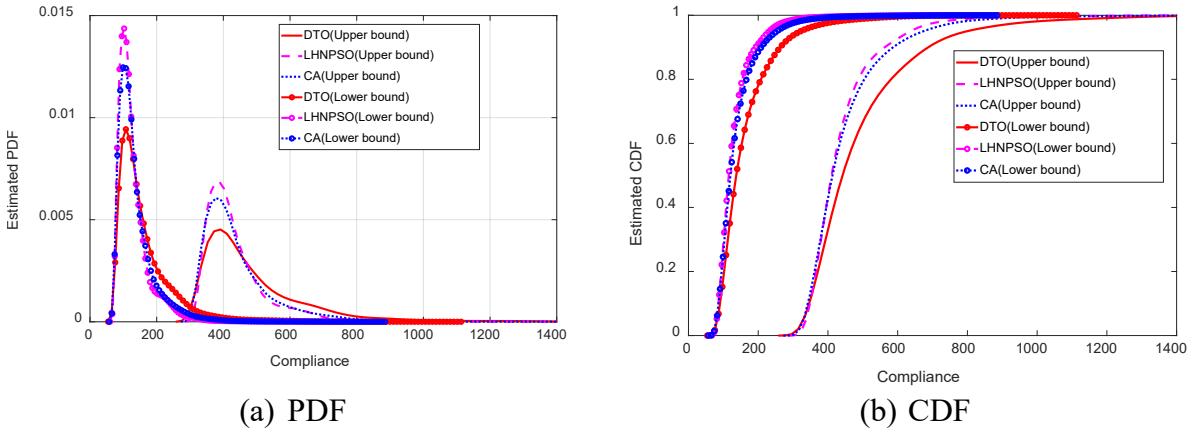

(a) PDF  (b) CDF

Figure 8 The upper bounds and lower bounds of the estimated PDF and CDF for DTO, LHNPSO and CA

Table 2 The comparison of final results and computational times of different methods

| Methods | J(c) Upper | J(c) Lower | μ(c) Upper | μ(c) Lower | σ(c) Upper | σ(c) Lower | Computational time |
|---|---|---|---|---|---|---|---|
| DTO | 551.991 | 192.494 | 488.441 | 161.223 | 63.550 | 31.271 | N/A |
| LHNPSO | 475.014 | 149.201 | 444.944 | 135.957 | 30.070 | 13.244 | 26.5 hrs |
| CA | 475.231 | 150.073 | 444.084 | 136.320 | 31.147 | 13.753 | 1.4 hrs |

6.1.4 *The influence of weight coefficients $w_1, w_2$ in Eq. (30)*

To consider the contribution of both the lower and upper bounds of the imprecise random parameters, two weighting coefficients $w_1$, $w_2$ are introduced. In this subsection, the influences of $w_1$, $w_2$ are investigated. The results of six different scenarios with $w_1$ and $w_2$ taking values from 0 to 1 with a 0.2 spacing are presented in Table 3. It can be seen that as $w_1$ increases and $w_2$ decreases, both the upper and lower bounds of the objective function value, mean and standard deviation of compliance gradually decrease. There are two possible reasons for this: Firstly, for different cases the parameters are monotonically changed. Secondly, as the contribution of $w_2$ increases, the influence of the lower bounds of the imprecise randomness increases. This can be clearly observed from the magnification of the peak in Figure 9, which depicts the upper bounds of the various cases.

Table 3 Comparison of the objective function value, mean and standard deviation of compliance for different weighting coefficients

| Parameter | Bound | Case 1 ($w_1$=1, $w_2$=0) | Case 2 ($w_1$=0.8, $w_2$=0.2) | Case 3 ($w_1$=0.6, $w_2$=0.4) | Case 4 ($w_1$=0.4, $w_2$=0.6) | Case 5 ($w_1$=0.2, $w_2$=0.8) | Case 6 ($w_1$=0, $w_2$=1) |
|---|---|---|---|---|---|---|---|
| J(c) | Upper | 475.231 | 474.892 | 473.655 | 466.618 | 449.321 | 437.679 |
| J(c) | Lower | 150.073 | 149.856 | 148.771 | 144.998 | 136.525 | 129.707 |
| μ(c) | Upper | 444.084 | 444.004 | 442.936 | 437.692 | 422.424 | 413.686 |
| μ(c) | Lower | 136.320 | 136.233 | 135.309 | 132.470 | 124.872 | 119.304 |
| σ(c) | Upper | 31.147 | 30.888 | 30.719 | 28.926 | 26.897 | 23.993 |
| σ(c) | Lower | 13.753 | 13.623 | 13.463 | 12.528 | 11.653 | 10.403 |

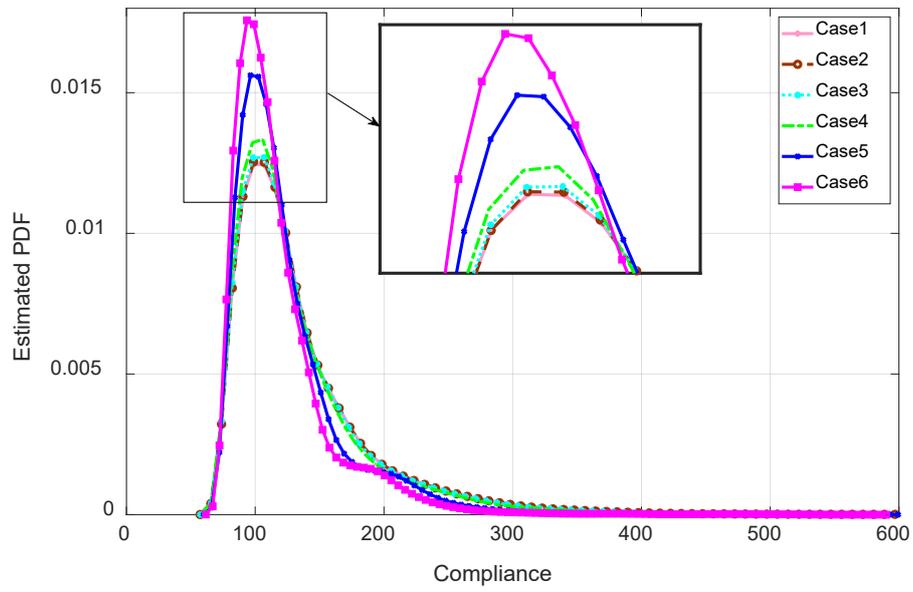

Figure 9 The upper bounds of estimated PDFs of different cases

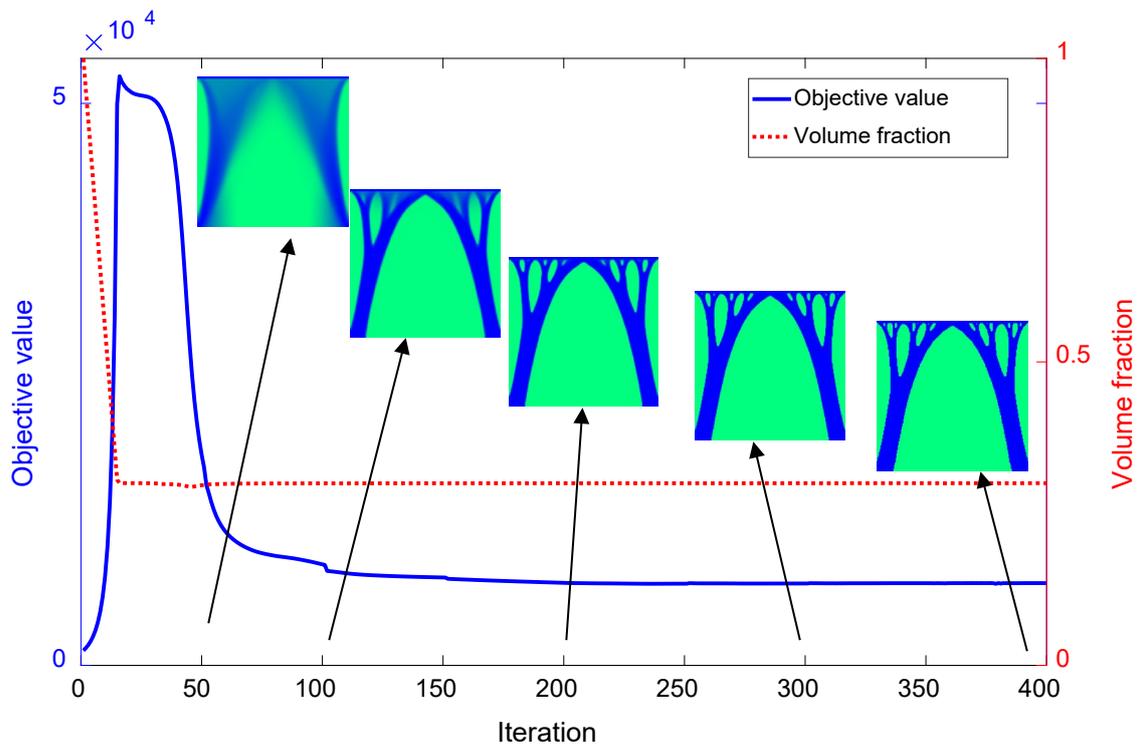

Figure 10 Iteration history of Case 1 ($w_1 = 1$; $w_2 = 0$)

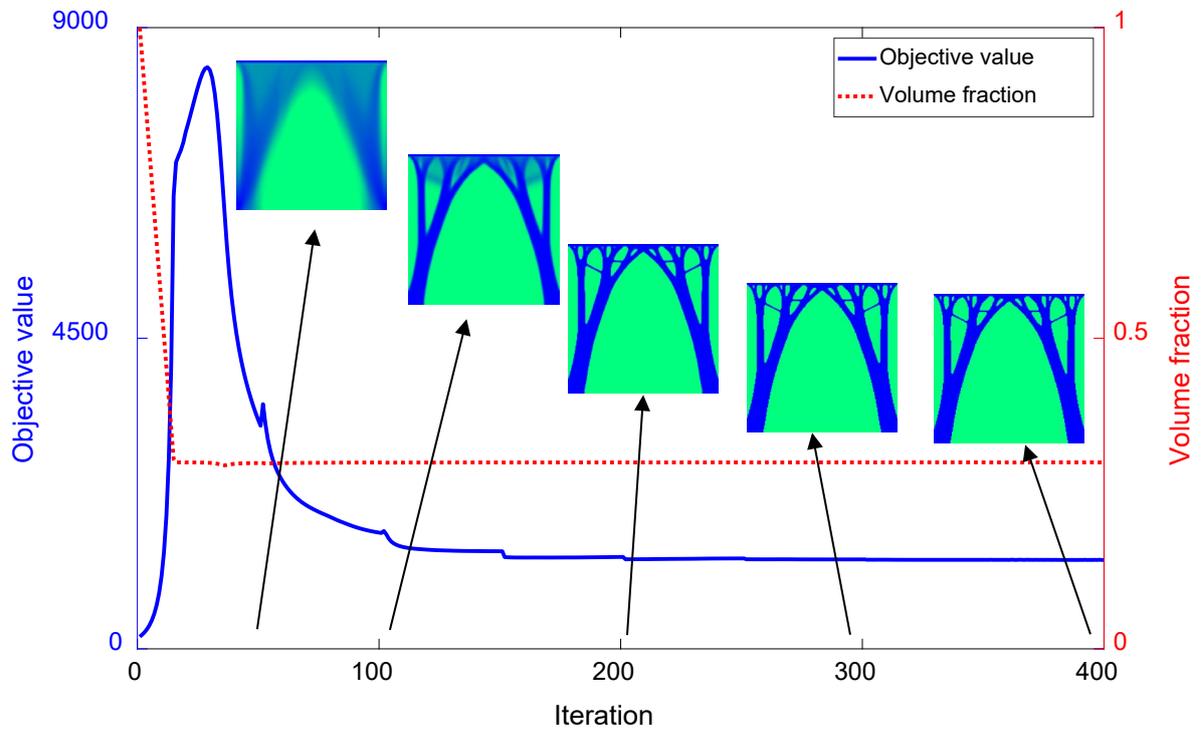

Figure 11 Iteration history of Case 6 ($w_1 = 0$; $w_2 = 1$)

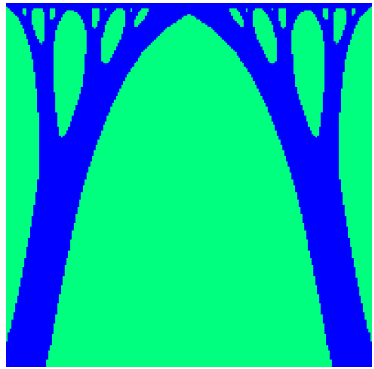
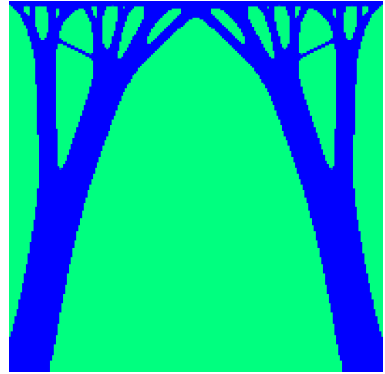

(a) Case 1 ($w_1=1, w_2=0$)     (b) Case 2 ($w_1=0.8, w_2=0.2$)

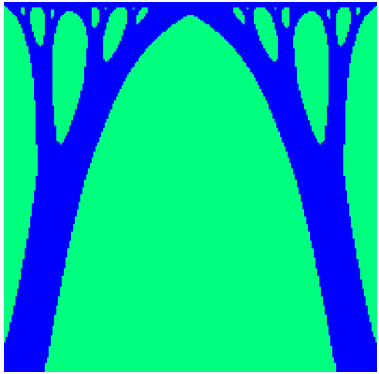
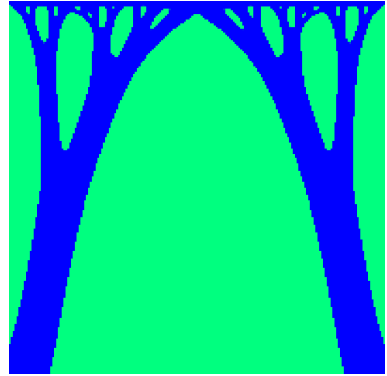

(c) Case 3 ($w_1=0.6, w_2=0.4$)     (d) Case 4 ($w_1=0.4, w_2=0.6$)

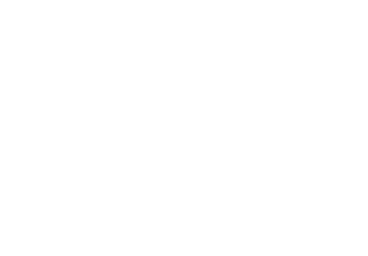
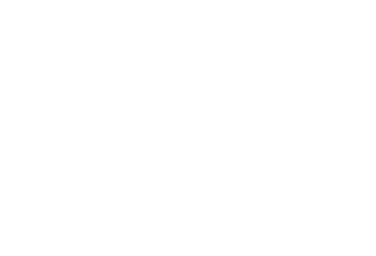

(e) Case 5 ($w_1=0.2, w_2=0.8$)     (f) Case 6 ($w_1=0, w_2=1$)

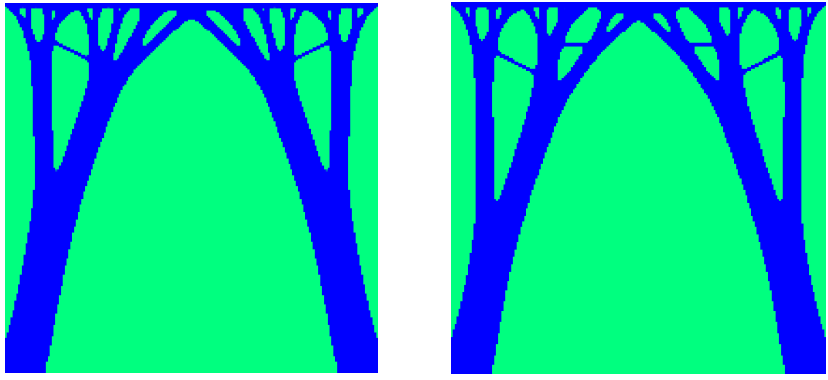

Figure 12 Robust design of Carrier plate with imprecise random field loads for different cases

For the sake of brevity, Figure 10 and Figure 11 show the iteration history of the objective function and volume fraction for Case 1 and Case 6, respectively. The structural volume decreases rapidly (the red dash line) and the objective value (the blue solid line) are satisfied after 100 iterations. Figure 12 presents the possible optimized designs of the Carrier plate with the imprecise random field. It should be noted that the objective function values as well as the mean and standard deviation of the compliance in each case are smaller than the DTO values, as shown in Table 2 and Table 3. Therefore, the proposed method provides improved performance in achieving a robust solution as well as options for topological layouts of a structure within a wide range of uncertainties.

6.2. *Numerical experiment 2: cantilever beam with uncertain tip forces*

To show the application of the proposed method, a study of optimization of periodic topologies under imprecise random field loads is presented here. Considering the periodicity of a structure in the given domain, for example, for a 2D structure, the design domain is divided into $X(k, l) = N_x \times N_y$ unit cells where $N_x$, $N_y$ represent the number of unit cells along the $x$ and $y$ directions, respectively. Each of the unit cells, can be further separated into several smaller elements $x(i, j) = n_x \times n_y$, as shown in Figure 13, where $n_x$ and $n_y$ denote the number of elements in the $x$ and $y$ directions, respectively. Therefore, the RTO problem for periodic structures can be defined as:

$$\begin{aligned}
&design \quad (x, X) \\
&\min \quad J(x, X) = J(\tilde{x}, \tilde{X}) \\
&s.t. \quad \begin{cases} K(\tilde{x})U = F \\ \tilde{X}(k,l) \geq \sum \tilde{x} \\ \sum \tilde{X} \leq \bar{V} \end{cases} \\
&x(i, j) = x((N_x - 1)n_x + i, j) \quad N_x = 1 \sim k \\
&X(k.l) = X(k + N_x - 1, l) \quad N_x = 1 \sim k
\end{aligned} \quad (54)$$

where $x(i,j)$ is an element design variable with $i$ and $j$ denoting the element number along the two axes and $X(k,l)$ is a cell design variable where $k$ and $l$ are the unit cell number along the two axes. $\tilde{x}, \tilde{X}$ represent the variables after filtering.

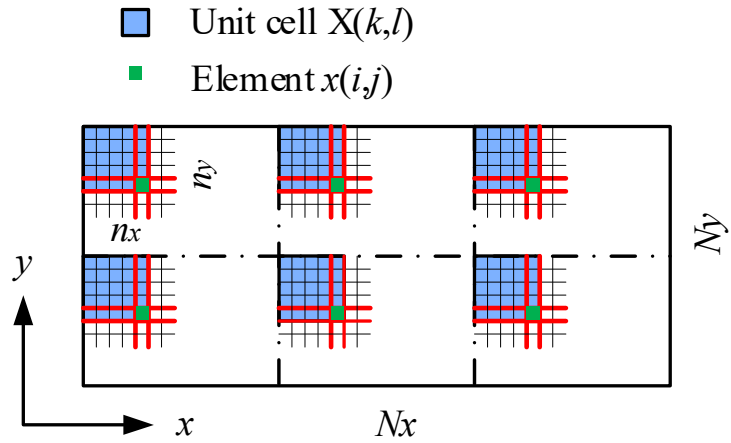

Figure 13 A periodic structure with unit cells

A cantilever beam with imprecise random field tip forces is considered first. The dimensions of the beam are 300×90 with a load applied over the first 40 elements on the left side, as shown in Figure 14. The structure is divided into $X(24, 6)$ unit cells each containing $x(15,15)$ elements. Users can adjust these parameters according to design specifications. Filter radii of 1.5 for the element and 5 for the unit cell are chosen for the improved Heaviside filtering process. The total structural volume fraction is set to 0.5. The mean and standard deviation of the load magnitude are -1 and 0.5, respectively. The correlation length of the distributed loads is 5 in this experiment. A 90% confidence interval is considered.

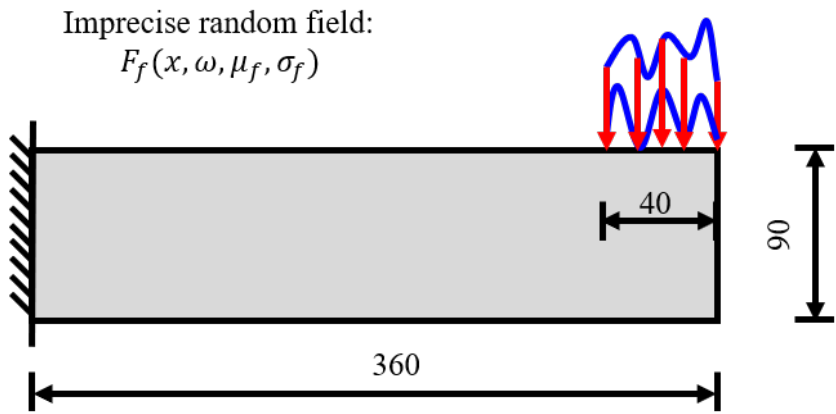

Figure 14 Cantilever beam with the imprecise random field tip forces

As discussed in subsection 6.1.4, the topologies resulting from RTO as well as those obtained using DTO are introduced in Table 4. As can be seen, those from DTO are very similar to those for the proposed method, although there are some minor differences. However, the statistical data (expected values and standard deviations) of the RTO solutions are smaller than those of the corresponding DTO results, revealing that the results from the RTO are more robust to imprecise randomness than the deterministic structure. Figure 15 presents the iteration history for the cantilever beam, with both the objective value and volume fraction illustrated. Three sharp peaks are seen to occur in the 22$^{nd}$, 32$^{nd}$ and 54$^{th}$ iterations in both the objective value and the volume fraction during the topology optimization. After that, the results converge to an optimal design at around the 120$^{th}$ loop.

Table 4 The comparison of optimal topologies between DTO and RTO

| Scenarios | Optimal design | Part | Statistical information |
|---|---|---|---|
| DTO | 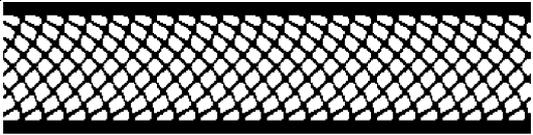 | 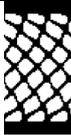 | $\mu(c)$= [597.176  972.647]<br>$\sigma(c)$= [256.929  456.228] |
| RTO<br>($w_1$=1, $w_2$=0) | 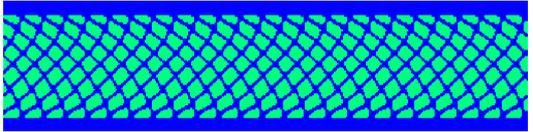 | 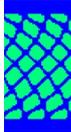 | $\mu(c)$=[580.267  945.108]<br>$\sigma(c)$=[249.653  443.307] |

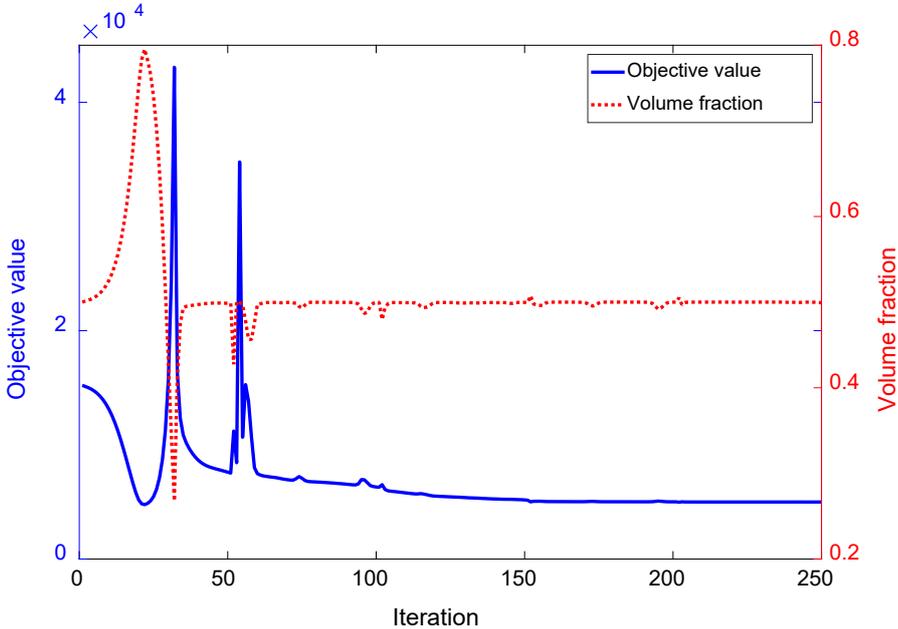

Figure 15 Iteration histories for the cantilever beam (Case 1)

The influence of $\beta$ in Eq. (11) on the optimal topologies is presented in Table 5. By increasing the value of $\beta$, the mean and standard deviation increase simultaneously and the junctions between intersecting braces become thicker. Increases in $\beta$ will therefore strengthen the constraint on variability.

Table 5 The influence of $\beta$ on optimal topologies

| Scenarios | Optimal design | Part | Statistical information |
|---|---|---|---|
| $\beta=1$ | 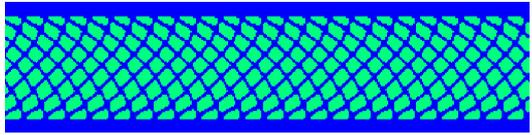 | 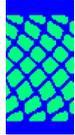 | $\mu(c)$=[580.267 945.108]<br>$\sigma(c)$=[249.653 443.307] |
| $\beta=2$ | 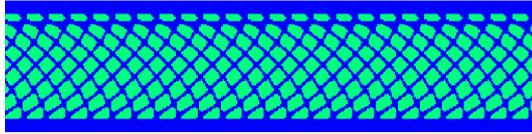 | 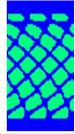 | $\mu(c)$=[ 583.229 949.930]<br>$\sigma(c)$=[ 250.924 445.566] |
| $\beta=3$ | 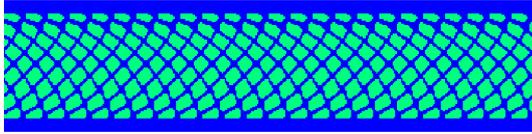 | 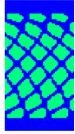 | $\mu(c)$=[ 588.590 958.664]<br>$\sigma(c)$=[ 253.231 449.661] |

6.3.  *Numerical experiment 3: Michell-type structure under imprecise random field loads*

This section presents a Michell-type structure with periodic topology under imprecise random field loads. The dimensions of the domain are 360×120, as shown in Figure 16. The structure is meshed using linear quadrilateral elements. The periodic structures is assumed to have $X$ (24, 8) unit cells each comprising $x$ (15,15) small elements. The filter radii of the elements and unit cells are same as in example 2. The mean and standard deviation of the load magnitude are -1 and 0.2, respectively.

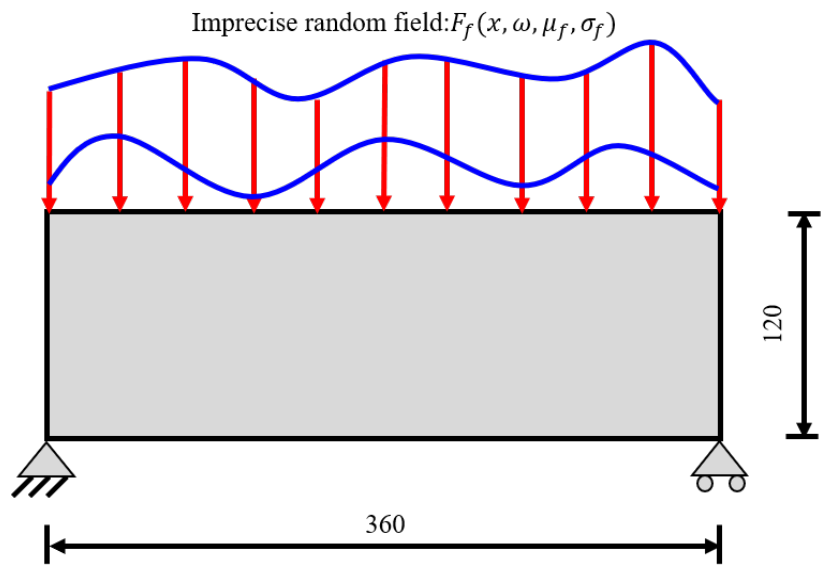

Figure 16 Michell-type structure with the imprecise random field loads

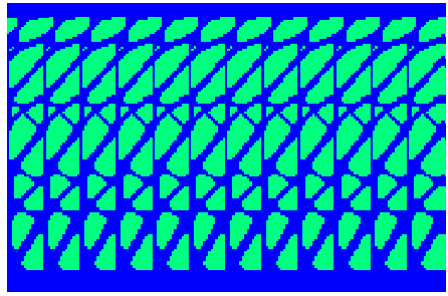

Figure 17 The final design of Case1 (1/2 structure)

To demonstrate the effectiveness of the proposed method, Figure 17 shows the resulting topology of the Michell-type structure with imprecise random field uncertainty and the corresponding iteration histories for the objective value and volume fraction are given in Figure 18. The following observations can be made from these two figures: Firstly, as the boundaries of the Michell-type structure are constrained at both ends, the appearance of the diagonal braces helps the structure withstand the uncertain force. Secondly, if we examine the structure from top to bottom, there are 4 layers over which the diagonal braces gradually become thicker. Moreover, connections between vertical bars and the diagonal braces develop in the middle area to resist local collapse.

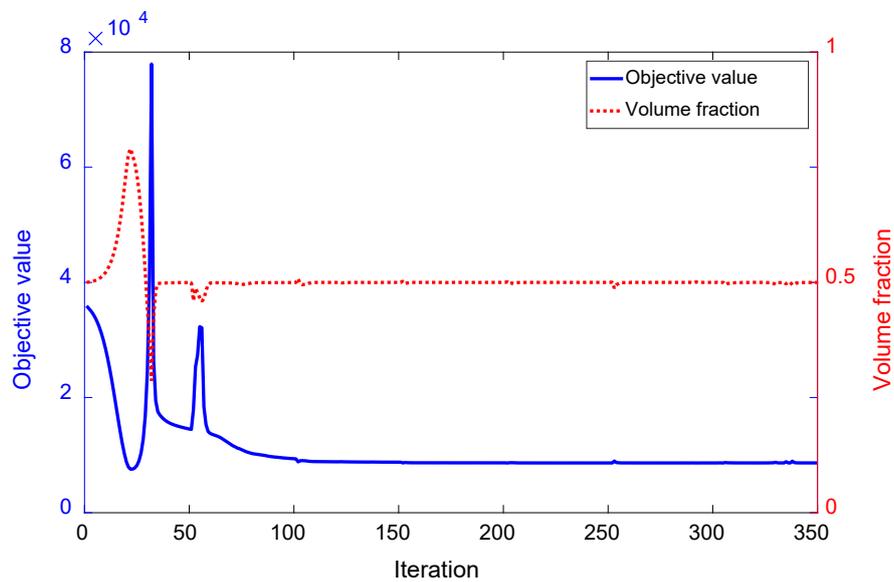

Figure 18 Iteration histories of Michell-type structure (Case 1)

Table 6 shows the comparison of optimal topologies between DTO and RTO. The RTO results are more robust and less sensitive to uncertain loads. The second significant difference between the RTO and DTO outcomes are that there are no connections between the vertical bars and diagonal braces in the DTO solutions. For different robust design cases, the differences are quite small although we can observe minor changes in the vertical bars, diagonal braces and the connections between them. This phenomenon is verified by Figure 19, which gives upper and lower bounds of the estimated PDFs and CDFs for different cases.

Table 6 The comparison of optimal topologies between DTO and RTO

| Scenarios | Optimal design | Statistical information |
|---|---|---|
| DTO | 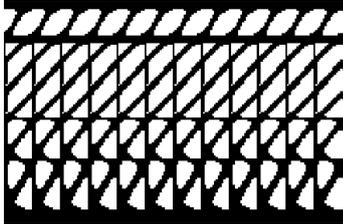 | $\mu(c)$=[ 1048.040 1283.287]<br>$\sigma(c)$=[ 282.897 435.088] |
| RTO<br>($w_1$=1, $w_2$=0) | 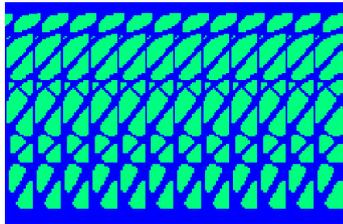 | $\mu(c)$=[1028.204 1258.880]<br>$\sigma(c)$=[277.514 426.804] |

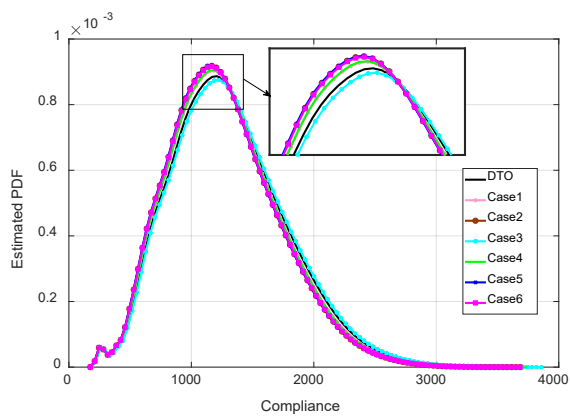 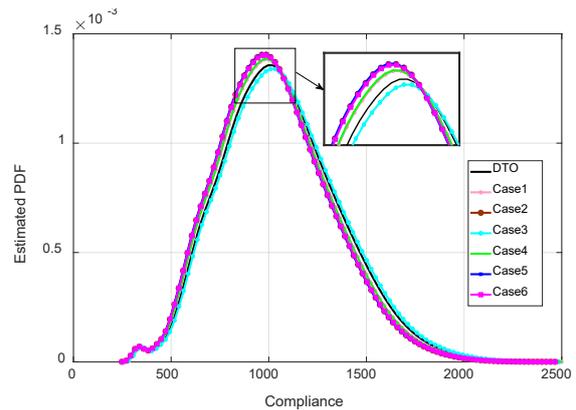

(a) The upper bounds of the estimated PDFs  (b) The lower bounds of the estimated PDFs

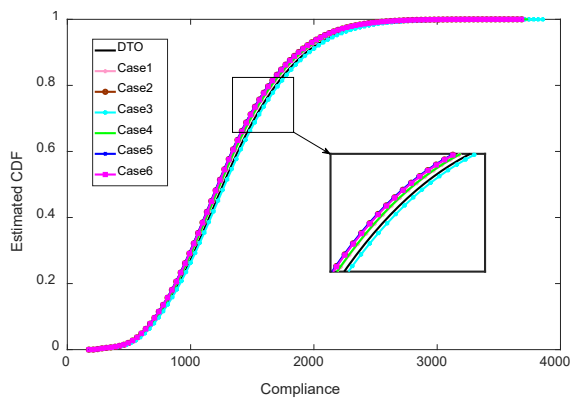 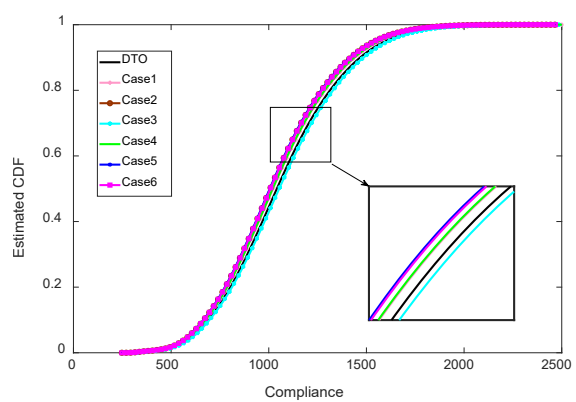

(c) The upper bounds of the estimated CDFs  (d) The lower bounds of the estimated CDFs

Figure 19 The upper lower bounds of the estimated PDFs and CDFs of the proposed method for different weightings

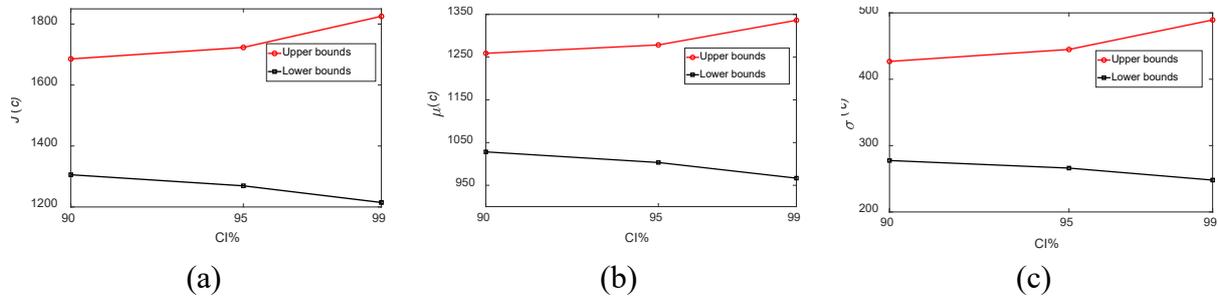

(a)                                     (b)                                     (c)

Figure 20 The influence of percentage confidence interval on (a) Objective values, (b) mean, (c) Standard deviation

Figure 20 illustrates the influence of three different confidence levels (CIs), 90%, 95% and 99% on the objective values, means and standard deviations of compliance. It is demonstrated that the increase in CI% widens the range of the upper and lower bounds as the statistical moments of the inputs increase.

## 7. Conclusion

A new computational method for RTO problems subject to imprecise random fields is presented in this study. Imprecise random loading is introduced into both continuum structures and periodic structures. Compared with the worst-case approach of RTO, the present method provides the upper and lower bounds of mean and standard deviation of compliance as well as the corresponding alternative topological layouts for various weightings. To consider the various confidence levels, the imprecise random fluctuation is illustrated by adopting a parameterized *p*-box feature and spectral description via the *K-L* expansion towards the loading domain. Based on the assumptions of the linear superposition method and the linear combination of orthogonal functions, an explicit mathematical expression of the statistical moments of structural compliance is rigorously proved. Then, an interval sensitivity analysis is derived by applying the OST method and the boundaries of each intermediate variable are efficiently searched for at every iteration by CA. The validity, accuracy and effectiveness of the proposed new method are comprehensively verified against the LHNPSO and QMCS methods. Results show excellent agreements among CA, LHNPSO and QMCS while CA obtains superior time-saving properties for a monotonic system. Finally, three different numerical examples with imprecise random field loads are presented to show the accuracy and feasibility of the study. The proposed method provides a more robust design with multiple topologies available under uncertain loads and is capable of investigating extreme upper and lower bounds with imprecise randomness.

**Acknowledgements**

The work described in the present paper is fully funded by "Engineering fellowships for growth: Materials by design for impact in aerospace engineering", EP/M002322/2 funded by EPSRC. The authors are grateful for the financial support.

**Declaration of Interests**

All authors declare that they do not have any conflict of interest in the work presented in this paper.